\documentclass[aps,preprint,prb]{revtex4}
\usepackage{graphicx}
\usepackage{epsfig}
\usepackage{amsmath}
\usepackage{graphics}
\begin{document}

\author{M. F. Gelin}
\author{D. S. Kosov}

\affiliation{Department of Chemistry and Biochemistry,
 University of Maryland, 
 College Park, 
 MD 20742 }

\title{Manifestation of  nonequilibrium initial conditions in molecular rotation: the generalized J-diffusion model}

\begin{abstract}

In order to adequately describe molecular rotation far from equilibrium,
we have generalized the J-diffusion model by allowing the
rotational relaxation rate to be angular momentum dependent. The calculated
nonequilibrium rotational correlation functions (CFs) are shown to
decay much slower than their equilibrium counterparts, and orientational
CFs of hot molecules exhibit coherent behavior, which persists for
several rotational periods. As distinct from the results of standard
theories, rotational and orientational CFs are found to dependent
strongly on the nonequilibrium preparation of the molecular ensemble. We predict
the Arrhenius energy dependence of rotational relaxation times and
violation of the Hubbard relations for orientational relaxation times.
The standard and generalized J-diffusion models
are shown to be almost indistinguishable under equilibrium conditions.
Far from equilibrium, their predictions may differ dramatically.

\end{abstract}
\maketitle

\section{Introduction}

There have been collected many evidences (due to
polarization time-resolved experiments, \cite{ruh93,wie93,wie96,zew96,hoc96,hoc97,hoc97a,voh98,voh99,voh99a,vol99,voh00,voh00a,nib01,bra03}
computer simulations, \cite{wil89,ger94,ben93,ben95,ben03,gri84}
and model studies \cite{gri84,gel00,gel01}) which indicate that rotational
and orientational relaxations of molecules in liquids under equilibrium and (highly)
nonequilibrium conditions differ significantly. The studies of nonequilibrium rotational relaxation
have recently culminated in papers. \cite{str06,str06a} As has been
demonstrated, relaxation of rotational energy of hot nonequilibrium
photofragments slows significantly with the increase of their initial
rotational temperature and differs dramatically from relaxation
of the equilibrium rotational energy correlation function, manifesting
thereby breakdown of the linear response description. Furthermore,
hot nonequilibrium molecules exhibit almost dissipation-free coherent
rotation, which persists for several rotational periods. Molecular
dynamic simulations have attributed this unusual behavior to a rapid
rearrangement of the local liquid structure: a hot solute pushes away
the neighboring solvent molecules and rotates almost freely until
it looses enough energy. \cite{str06,str06a} Another way of thinking
about that is in terms of the angular momentum dependent rotational
friction: highly rotationally excited molecules experience lower
friction than their thermally equilibrated counterparts. \cite{str06a,kos06d}

These nonequilibrium relaxations are not reproduced by conventional
theories of (equilibrium) molecular rotation,
which are based upon  master equations with
constant relaxation rates (the extended diffusion models, \cite{gor66,rid69,McCl77,BurTe}
the Keilson-Storer model \cite{sack,BurTe,kos06}) or rotational
Fokker-Planck equations with constant frictions. \cite{hub72,McCo,mor82,McCl87}
Physically, this is not surprising. If we consider relaxation under
equilibrium conditions, then all the relevant energies are of the
order of $k_{B}T$ ($k_{B}$ being the Boltzmann constant and $T$
being the temperature of the bath). So, putting aside non-Markovian
effects, we can always introduce a certain effective constant relaxation rate.
If we consider relaxation under nonequilibrium conditions, when initial
energies are much higher than $k_{B}T$, such a description is no
longer applicable, and the angular momentum dependence of relaxation rates
must be explicitly taken into consideration (see Ref. \cite{str06a,kos06d} for
more detailed argumentation and discussion). 

There exist two major and complimentary approaches to
rotational relaxation, via the Fokker-Planck equations \cite{hub72,McCo,mor82,McCl87}
and extended diffusion models. \cite{gor66,rid69,McCl77,BurTe} In
our previous paper, \cite{kos06d} we have demonstrated how the angular-momentum
dependent friction can be incorporated into the rotational Fokker-Planck
equation. The present paper deals with a similar generalization of
the J-diffusion model.  First, we investigate
how the angular momentum dependent rates manifest themselves in relaxations
of initial nonequilibrium distributions and in behaviors of rotational
and orientational correlation functions (CFs). Second, we develop
a reliable tool for the description of nonequilibrium molecular
rotation, which can be used i.e. for the interpretation of photofragment
anisotropy decays. \cite{ruh93,wie93,wie96,zew96,hoc96,hoc97,hoc97a,voh98,voh99,voh99a,vol99,voh00,voh00a,nib01,bra03}
Third, we demonstrate how the use of standard ''equilibrium'' approaches
may lead to quantitatively and even qualitatively wrong results,
if they are applied to describe nonequilibrium molecular rotation. 

Photodissociation $\mathrm{ICN}+h\nu\rightarrow\mathrm{CN+I}$ (which
was studied experimentally in the gas-phase \cite{zew94,zew01,wit85}
and in the condensed phase \cite{zew96,bra03}, and also investigated via computer
simulations \cite{wil89,ger94,ben95,ben03,str06,str06a}) is chosen 
in the present article as a prototype process which produces highly
nonequilibrium hot photofragments. The results of the molecular dynamics simulations
\cite{str06,str06a} are used to obtain realistic values of the model
parameters and to test our theoretical predictions. 

The structure of our paper is as follows. The generalized J-diffusion
model, which accounts for the angular momentum dependence of the relaxation
rate, is developed in Sec. 2. A convenient explicit expression for
the initial nonequilibrium angular momentum distribution is derived
in Section 3. Relaxation of this nonequilibrium distribution, as well
as dynamics of the angular momentum and energy CFs, are investigated
in Section 4. Orientational relaxation is studied in Section 5. The
Appendixes describe the methods of calculation of rotational
CFs (A), their relaxation times (B), and orientational CFs (C). 

The reduced variables are used throughout the article: time,
angular momentum and energy are measured in units of $\sqrt{I/(k_{B}T})$,
$\sqrt{Ik_{B}T}$ and $k_{B}T$, respectively. Here $k_{B}$ is the
Boltzmann constant, $T$ is the temperature of the equilibrium bath,
and $I$ is the moment of inertia of the photofragment,
so that $\tau_{r}=\sqrt{I/(k_{B}T})$ is the averaged period of its
free rotation. For CN at $300$ K, $\tau_{r}=$$0.2$ ps. All the
Laplace-transformed operators are denoted by tilde, viz. \begin{equation}
\tilde{f}(s)\equiv\int_{0}^{\infty}dt\exp\{-st\} f(t)\,\,\mathrm{for}\,\,\forall\, f(t).\label{Lap}\end{equation}

\section{Generalized J-diffusion model}

Restricting our consideration to linear molecules and neglecting the
short-time non-Markovian effects, we start from the  rotational
master equation

\begin{equation}
\partial_{t}\rho(\mathbf{J},\mathbf{\Omega},t)=-i\mathbf{J}\hat{\mathbf{L}}\rho(\mathbf{J},\mathbf{\Omega},t)-\int d\mathbf{J}'T(\mathbf{J}'|\mathbf{J})\rho(\mathbf{J},\mathbf{\Omega},t)+\int d\mathbf{J}'T(\mathbf{J}|\mathbf{J}')\rho(\mathbf{J}',\mathbf{\Omega},t).\label{kin1}\end{equation}
 Here $\rho(\mathbf{J},\mathbf{\Omega},t)$ is the probability density
function, $\mathbf{J}$ is the (two-dimensional) angular momentum
of the linear top in its molecular frame, $\mathbf{\Omega}$ are
the Euler angles which specify orientation of the molecular frame
with respect to the laboratory one. The first term on the right-hand
side of Eq. (\ref{kin1}) is the free-rotor Liouville operator, which
describes the angular momentum driven reorientation, $\hat{\mathbf{L}}$
being the angular momentum operator in the molecular frame. The last
two terms are responsible for the bath-induced relaxation. Since our molecular ensemble
is isotropic, the
kernel of the dissipation operator, $T(\mathbf{J}|\mathbf{J}')$,
is $\mathbf{\Omega}$-independent. It must obey the detailed balance
\begin{equation}
T(\mathbf{J}|\mathbf{J}')\rho_{B}(\mathbf{J}')=T(\mathbf{J}'|\mathbf{J})\rho_{B}(\mathbf{J}),\label{DetBal}\end{equation}
which is responsible for bringing the system under study to the equilibrium
rotational Boltzmann distribution\begin{equation}
\rho_{B}(\mathbf{J})=(2\pi)^{-1}\exp\{-\mathbf{J}^{2}/2\}.\label{Boltz}\end{equation}
 Normalization of Eq. (\ref{kin1}),\begin{equation}
\int d\mathbf{J}d\mathbf{\Omega}\rho(\mathbf{J},\mathbf{\Omega},t)\equiv1,\label{norm}\end{equation}
 which insures conservation of probability, is accounted for automatically.

By selecting a particular form of the relaxation kernel $T(\mathbf{J}'|\mathbf{J})$,
one can recover many models of rotational relaxation available in
the literature: the extended diffusion models, \cite{gor66,rid69,McCl77,BurTe}
the rotational Fokker-Planck equation, \cite{hub72,McCo,mor82,McCl87}
the Keilson-Storer model. \cite{sack,BurTe,kos06} All these models,
however, yield $\int d\mathbf{J}'T(\mathbf{J}'|\mathbf{J})=\nu=\textrm{const}$,
that is the collision rate is angular momentum independent. For example,
the most popular model of rotational relaxation, the J-diffusion model,
corresponds to the choice \begin{equation}
T(\mathbf{J}|\mathbf{J}')=\nu\rho_{B}(\mathbf{J}).\label{JD}\end{equation}
We cannot straightforwardly generalize the J-diffusion model by replacing
the rate $\nu$ with its $J$-dependent counterpart $\nu(J)$, since
this procedure would violate normalization (and therefore conservation)
of the probability density (Eq. (\ref{norm})). To get a correct description,
we assume that the relaxation kernel can be represented in the factorized
Gaussian form 

\begin{equation}
T(\mathbf{J}|\mathbf{J}')=\exp\{-a\mathbf{J}^{2}/2\}\exp\{-b\mathbf{J'}^{2}/2\} c,\label{GJDkern}\end{equation}
$a,\, b$, and $c$ being numerical constants to be specified. To
satisfy the detailed balance (\ref{DetBal}), we must choose \begin{equation}
a=1+b.\label{h}\end{equation}
Then the generalized J-diffusion relaxation kernel takes the form
\begin{equation}
T(\mathbf{J}|\mathbf{J}')=\nu\frac{1+b}{2\pi}\exp\{-(1+b)\mathbf{J}^{2}/2\}\exp\{-b\mathbf{J'}^{2}/2\}.\label{GJD}\end{equation}
Here the rate $\nu$ governs the dissipation strength and the constant 

\begin{equation}
c=\nu\frac{1+b}{2\pi}\label{c}\end{equation}
is chosen to insure that \begin{equation}
\int d\mathbf{J}'T(\mathbf{J}'|\mathbf{J})\equiv\nu z(J),\,\,\, z(J)\equiv\exp\{-bJ^{2}/2\}.\label{z(j)}\end{equation}
Thus, $\nu z(J)$ can be regarded as $J$-dependent collision frequency.
\cite{FootG} 

Eq. (\ref{GJDkern}) looks as a natural generalization of the standard
J-diffusion kernel (\ref{JD}), which still preserves its Gaussian
and factorized form. If $b$ is small (it is indeed small, see bellow),
then the kernel (\ref{GJDkern}) can be regarded as the first order
correction to its J-diffusion counterpart (\ref{JD}). Furthermore,
any relaxation kernel $T(\mathbf{J}'|\mathbf{J})$ can be expanded
in series on Gaussians,\begin{equation}
T(\mathbf{J}|\mathbf{J}')=\sum_{b}\nu_{b}\frac{1+b}{2\pi}\exp\{-(1+b)\mathbf{J}^{2}/2\}\exp\{-b\mathbf{J'}^{2}/2\},\label{Gauss}\end{equation}
$\nu_{b}$ being certain expansion coefficients. \cite{foot3} Thus, Eq. (\ref{GJDkern})
retains the first term in this expansion, and
can further be refined by taking more terms, if necessary. The use
of Eq. (\ref{GJDkern}) can be justified \textit{a posteriori,} since
the predictions of the generalized J-diffusion model in Secs. 4 and
5 sustain comparison with the results of molecular dynamics simulations
performed in Refs. \cite{str06,str06a}

Plugging the kernel (\ref{GJDkern}) into Eq. (\ref{kin1}) we obtain
our generalized J-diffusion master equation: \[
\partial_{t}\rho(\mathbf{J},\mathbf{\Omega},t)=-i\mathbf{J}\hat{\mathbf{L}}\rho(\mathbf{J},\mathbf{\Omega},t)-\nu z(J)\rho(\mathbf{J},\mathbf{\Omega},t)\]
\begin{equation}
+\nu(1+b)z(J)\rho_{B}(\mathbf{J})\int d\mathbf{J}'z(J')\rho(\mathbf{J}',\mathbf{\Omega},t).\label{kin2}\end{equation}
 This is the equation which will be studied in the subsequent Sections.
When \textbf{$b=0$} ($z=1$), it reduces to the standard J-diffusion
model. 

Before embarking at particular calculations, it is useful to estimate plausible 
values of the parameters $\nu$ and $b$. Since $\nu$ describes the
overall dissipation strength and is similar to its counterpart in
the standard J-diffusion model, there are no intrinsic limitations
on the value of this parameter. It is small ($<1$) for rarefied gases
and large ($>1$) for liquids and solutions. As to the value of $b$,
it is expected to be small ($\ll1$), since the effects due to the
$J$-dependence of relaxation rates do not manifest themselves at
room temperatures under equilibrium conditions. On the other hand,
if the initial nonequilibrium energy of the photofragment $E_{\Delta}$ is much larger than $k_{B}T$,
 then the product $b(E_{\Delta}/k_{B}T)$ is
not necessarily small. It is in this case the effects due to the $J$-dependence
of relaxation rates become important. Note that the parameters $\nu$
and $b$ must be independent of initial conditions. However, as we shall
see in Secs. 4 and 5, Eq. (\ref{kin2}) predicts that nonequilibrium
responses, i.e. the ensuing rotational and orientational CFs, depend
strongly upon initial conditions.

\section{Nonequilibrium initial distribution}

Normally, master equations like (\ref{kin2}) are used to calculate
various rotational and orientational CFs under equilibrium conditions,
assuming that $\rho(\mathbf{J},\mathbf{\Omega},0)=(2\pi)^{-3}\rho_{B}(\mathbf{J})$.
We wish to study rotational relaxation 
under nonequilibrium initial conditions \begin{equation}
\rho(\mathbf{J},\mathbf{\Omega},0)=(2\pi)^{-3}\rho_{ne}(\mathbf{J}).\label{ne}\end{equation}

To get an explicit expression for $\rho_{ne}(\mathbf{J})$, we adopt
a model developed in Ref. \cite{gel02} Let us consider the photoreaction
$A+h\nu\rightarrow B+\mathrm{products}$. We assume that the photofragmentation
is instantaneous on the time scale of molecular rotation. This assumption
holds true for most of small \cite{wil89,ger94,ben93,ben95,ben03,zew94,zew01}
and polyatomic \cite{gel99} molecules, for which the photofragmentation
time can be be estimated by few hundreds of femtoseconds. \cite{FootA}
Having accepted the assumption about an instantaneous (impulsive)
photodecomposition, we immediately conclude that dissociation produces
a nonequilibrium distribution over the angular momenta of photoproducts.
Indeed, there exist two major sources of rotational excitation
of fragments. These are the parent molecule rotation and the applied
torque. Therefore, the angular momenta of the parent ($\mathbf{J}_{A}$)
and product ($\mathbf{J}_{B}$) molecules are connected through the
formula \cite{zew94,gel99}\begin{equation}
J_{B,\alpha}=\sum_{\alpha=x,y,z}I_{B,\alpha}R_{\alpha\beta}(\mathbf{\Xi})I_{A,\beta}^{-1}J_{A,\beta}+\Delta_{\alpha}.\label{Ja&b}\end{equation}
The first term describes mapping of the parent molecule rotation
into that of the product, and the angular momentum $\mathbf{\Delta}$
originates from the impulsive torque arising due to the rupture of
chemical bond(s) of the parent. The small Latin indexes label
the Cartesian components of the corresponding vectors and tensors,
$I_{A,\alpha}$ and $I_{B,\alpha}$ are the main moments of inertia
of the species A and B, $R_{\alpha\beta}(\mathbf{\Xi})$ is the matrix
of rotation from the frame of the main moments of inertia of the product
to that of the parent, and $\mathbf{\Xi}$ are the pertinent Euler
angles. 

Let the parent molecule A be a planar asymmetric top (i.e., a triatomic
molecule, $I_{A,x}=I_{A,y}+I_{A,z}$) and the product B be a linear
rotor ($I_{B,x}=I_{B,y}=I_{B},\,\, I_{B,z}=0$). We then define the
molecular frames of the parent ($x_{A},\, y_{A},\, z_{A}$) and product
($x_{B},\, y_{B},\, z_{B}$) molecules in such a way that the axes
$x_{A}||x_{B}$ are perpendicular to the plane of the parent molecule,
and $z_{B}$-axis coincides with that of the linear fragment (see
Fig. 1). In this case \begin{equation}
R(\mathbf{\Xi})=\left(\begin{array}{ccc}
1 & 0 & 0\\
0 & \cos(\psi_{B}) & \sin(\psi_{B})\\
0 & -\sin(\psi_{B}) & \cos(\psi_{B})\end{array}\right),\label{R}\end{equation}
$\psi_{B}=\angle(y_{A}y_{B})=\angle(z_{A}z_{B})$. It is natural to
surmise that the applied torque is perpendicular to the 
plane of the parent molecule, $\mathbf{\Delta}||x_{B}$. Then, assuming that the parent molecules
have a Boltzmann equilibrium distribution at the bath temperature
$T$, we find that the linear photoproducts are distributed according
to \cite{gel02} \begin{equation}
\rho_{ne}(\mathbf{J})=\frac{\xi\eta}{2\pi^{2}}\int_{0}^{\pi}d\theta\exp\left\{ -\frac{\eta}{2}\left([J\cos\theta+\Delta]^{2}+\xi^{2}J^{2}\sin^{2}\theta\right)\right\} \label{neMy}\end{equation}
(hereafter, the subscript B, which denotes the angular momentum and
moment of inertia of fragment B, is omitted and $\Delta$ is the magnitude
of the impulsive angular momentum $\mathbf{\Delta}$).
We have introduced the parameters \cite{foot2}
\begin{equation}
\eta=\frac{I_{A,x}}{I},\,\,\,\xi^{2}=\frac{1}{I_{A,x}\{\cos^{2}(\psi_{B})/I_{A,y}+\sin^{2}(\psi_{B})/I_{A,z}\}}.\label{parNe}\end{equation}

Eq. (\ref{parNe}) is a desirable nonequilibrium distribution, which
is employed in all the subsequent calculations. The entire information
about dissociation is contained in the parameters $\Delta$, $\eta$,
and $\xi$. The first of them is responsible for the applied torque
and can be considered as dynamical. The last two parameters control
the parent-product rotational energy transfer. They are determined
by the geometry of the dissociating molecule. Note that $0\leq\xi\leq1$.
When $\xi\approx0$, then the parent molecule is highly prolate ($I_{A,z}\ll I_{A,x},\, I_{A,y}$)
and the linear fragment is \char`\"{}attached\char`\"{} perpendicularly
to the parent molecule axis ($\psi_{B}\approx\pi/2$). If $\xi\approx1$,
then the parent molecule is also considerably prolate, but the linear
fragment is \char`\"{}attached\char`\"{} parallel to the parent molecule
axis ($\psi_{B}\approx0$). If the parent molecule is a symmetric
top ($I_{A,x}=I_{A,y}$), then $\xi=1$. The quantities $\eta$
and $\xi$ can be regarded as known for a particular photoreaction,
since the angle $\psi_{B}$ and the main moments of inertia are fixed
for specific molecules A and B. The value of $\Delta$ can be
estimated through the consideration of the energy partitioning in
the course of the photofragmentation. \cite{zew94} 

Distribution (\ref{parNe}) is convenient for the further use, since
it covers many different scenario of the photofragmentation. It reduces
to the Boltzmann equilibrium distribution (\ref{Boltz}) if $\Delta=0$
and $\eta=$$\xi=1$. On the other hand, we get the delta-distribution
\begin{equation}
\rho_{ne}(\mathbf{J})=\frac{1}{2\pi}\frac{\delta(J-\Delta)}{J}\label{DeltaD}\end{equation}
 in the limit $\eta\rightarrow\infty$. 

The quantum version of distribution (\ref{neMy}) is immediately written
down after the {}``quantization'' of the angular momentum: \cite{gel02}
\begin{equation}
\rho_{ne}^{Q}(j)=Z_{Q}^{-1}\chi(j),\,\, Z_{Q}=\sum_{j=0}^{\infty}\chi(j),\label{neMyQ}\end{equation}
\[
\chi(j)=(2j+1)\int_{0}^{\pi}d\theta\exp\left\{ -\frac{\eta_{Q}}{2}\left([\overline{j}\cos\theta+\delta]^{2}+\xi^{2}\overline{j}^{2}\sin^{2}\theta\right)\right\} .\]
Here $j$ is the rotational quantum number, \begin{equation}
\overline{j}=\sqrt{j(j+1)},\,\,\,\delta=\Delta/\zeta,\,\,\,\eta_{Q}=\zeta^{2}\eta,\,\,\,\zeta\equiv\hbar/\sqrt{Ik_{B}T}.\label{parNeQ}\end{equation}
Eq. (\ref{neMyQ}) allows us to simulate a rotational distribution
of any width centered at any value of $j$. Several characteristic
shapes of $\rho_{ne}^{Q}(j)$ are presented in Fig. 2. If $\xi=1$,
the distribution is symmetric. If $\xi<1$ (for triatomics, this corresponds
to dissociation from a bent configuration) the distribution becomes
asymmetric, with a long-$j$ bias. 

Let us now apply the above machinery to photodissociation $\mathrm{ICN}+h\nu\rightarrow\mathrm{CN+I}$.
This reaction proceeds via two channels, producing low energy and
high energy photofragments. The quantum distribution $\rho_{ne}^{Q}(j)$
for nascent CN-fragments has been measured in. \cite{wit85} A comparison
of the {}``full line'' distribution in Fig. 2 with that reported
in \cite{wit85} reveals that the former reproduces qualitatively
the experimental results for hot photofragments. For CN at $300$
K, $\zeta=0.135$. Therefore, the {}``realistic values'' of the
dimensionless parameters $\eta=\eta_{Q}/\zeta^{2}=1.1$, $\Delta=\delta\zeta=5.4$
will be used in our illustrative calculations. \cite{FootC}

\section{Rotational relaxation}

In this section, we study evolution of the quantities which depend
on the angular momentum $\mathbf{J}$ but are independent of the Euler
angles $\mathbf{\Omega}$. We thus can integrate Eq. (\ref{kin2})
over $\mathbf{\Omega}$ and arrive at the reduced master equation\begin{equation}
\partial_{t}\rho(\mathbf{J},t)=-\nu z(J)\rho(\mathbf{J},t)+\nu(1+b)z(J)\rho_{B}(J)\int d\mathbf{J}'z(J')\rho(\mathbf{J}',t).\label{kinJ}\end{equation}
This equation can be used, for example, to calculate the angular momentum
CF \begin{equation}
C_{J}(t)=\frac{\left\langle \mathbf{JJ}(t)\right\rangle _{ne}}{\left\langle \mathbf{J}^{2}\right\rangle _{ne}},\label{CJ}\end{equation}
the averaged rotational energy\begin{equation}
C_{S}(t)=\frac{\left\langle E(t)\right\rangle _{ne}-\left\langle E\right\rangle _{B}}{\left\langle E\right\rangle _{ne}-\left\langle E\right\rangle _{B}},\label{Sne}\end{equation}
 and the rotational energy CF \begin{equation}
C_{E}(t)=\frac{\left\langle EE(t)\right\rangle _{ne}-\left\langle E\right\rangle _{B}\left\langle E\right\rangle _{ne}}{\left\langle E^{2}\right\rangle _{ne}-\left\langle E\right\rangle _{B}\left\langle E\right\rangle _{ne}}.\label{CBE}\end{equation}
 Here \begin{equation}
E\equiv\mathbf{J}^{2}/2\label{E&J}\end{equation}
and we use the notation $\left\langle ...\right\rangle _{a}=\int d\mathbf{J}\rho_{a}(\mathbf{J})...$,
$a=B,\, ne$. The corresponding integral relaxation times are determined
as

\begin{equation}
\tau_{a}=\int_{0}^{\infty}dtC_{a}(t),\,\,\,\,\, a=J,\, E,\, S.\label{Tau}\end{equation}

As is shown in Appendix A, the Laplace images of the probability density
$\rho(\mathbf{J},t)$, as well as of CFs (\ref{CJ})-(\ref{CBE})
can be given in quadratures (Eqs. (\ref{Ronet}), (\ref{Cab1}), and
(\ref{CabNewSF})) for any initial nonequilibrium distribution $\rho_{ne}(\mathbf{J})$.
The inversion of the Laplace transforms into the time domain can elementary
be performed numerically. Furthermore, we derive elucidating analytical expressions in
several important particular cases, which help us to grasp essential
features of rotational relaxation under nonequilibrium conditions. 

Let us assume that the nonequilibrium distribution is given by Eq.
(\ref{DeltaD}), that is the photoproducts are produced with a fixed
magnitude of the angular momentum $\Delta$. Such a situation corresponds
to the procedure of the (microcanonical) preparation of hot molecules 
in simulations reported in \cite{str06,str06a}.
We thus define the nonequilibrium photofragment energy \begin{equation}
E_{\Delta}=\Delta^{2}/2.\label{Edel}\end{equation}
We assume, in addition, that $b$ is small ($b\ll\eta,\,\eta\xi^{2}$)
but the product $bE_{\Delta}$ can take any value. Then Eq. (\ref{TauD})
predicts that the CFs (\ref{CJ})-(\ref{CBE}) are all the same and
exponential, \begin{equation}
C_{a}(t)=\exp\{-\nu z(\Delta)t\},\,\,\,\,\, a=J,\, E,\, S.\label{TauD1}\end{equation}
The $J$-dependent rates (\ref {z(j)}) are seen to slow rotational relaxation.
The corresponding relaxation times (\ref{Tau}) read \begin{equation}
\tau_{a}=\frac{1}{\nu z(\Delta)}\equiv\frac{1}{\nu}\exp\left\{ \frac{bE_{\Delta}}{k_{B}T}\right\} .\label{TauAll}\end{equation}
The last term in this expression is written in dimensional units,
to emphasize that the generalized J-diffusion model predicts the Arrhenius
energy dependence of the rotational relaxation times. Eq. (\ref{TauAll})
corroborates the finding \cite{str06,str06a,kos06d} that there exist
characteristic {}``nonequilibrium energy'' $\overline{E}_{\Delta}$
and {}``nonequilibrium temperature'' $\overline{T}_{\Delta}=\overline{E}_{\Delta}/k_{B}$,
\begin{equation}
\overline{E}_{\Delta}\sim k_{B}T/b,\,\,\,\overline{T}_{\Delta}\sim T/b\label{Char}\end{equation}
starting from which the $J$-dependence of rotational relaxation becomes
important. If the characteristic nonequilibrium energy $E_{\Delta}$
is much larger than $k_{B}T$, then the relaxation times for hot photofragments
can substantially be longer than their equilibrium counterparts $\tau_{a}=1/\nu$,
despite $b$ is small. The standard J-diffusion model, on the contrary,
predicts that any rotational CF (\ref{Cab}) decays exponentially
with the rate $\nu$, irrespective of particular forms of $A(\mathbf{J})$,
$B(\mathbf{J})$, and $\rho_{ne}(\mathbf{J})$ (see Eq. (\ref{JDcf})).
It is thus inadequate under highly nonequilibrium conditions,
when $bE_{\Delta}>1$. 

The Arrhenius energy dependence of the rotational relaxation times
correlates with the results of Tao and Stratt, \cite{str06,str06a}
who simulated CFs $C_{S}(t)$ (\ref{Sne}) for different initial nonequilibrium
temperatures  $\overline{T}_{\Delta}$. They obtained the following
values: (a) $\tau_{S}=5.29$ for $\overline{T}_{\Delta}=1917$ K,
(b) $\tau_{S}=7.73$ for $\overline{T}_{\Delta}=2395$ K, (c) $\tau_{S}=11.25$
for $\overline{T}_{\Delta}=2875$ K (see also Ref. \cite{kos06d}). 
We can take any two sets
of $\tau_{S}$ and $\overline{T}_{\Delta}$ to calculate $\nu$ and
$b$ by using Eq. (\ref{TauAll}). Doing so for three different pairings
of the sets, we have got remarkably consistent results: $\nu=0.87$,
$b=0.095$ for sets a\&b, $\nu=0.84$, $b=0.094$ for sets b\&c, $\nu=0.86$,
$b=0.095$ for sets c\&a. If we recall that Eqs. (\ref{TauD1}) and
(\ref{TauAll}) are approximate and valid in the limit of $b\ll1$,
$\eta\gg1$, the agreement becomes even more encouraging. Thus the
quantity $z(\Delta)$ is close to $1$ ($0.91$) at equilibrium ($\overline{T}_{\Delta}=120$
K), $0.22$ at $\overline{T}_{\Delta}=1917$ K , $0.15$ at $\overline{T}_{\Delta}=2395$
K, and $0.11$ at $\overline{T}_{\Delta}=2875$. So, rotational relaxation
at $\overline{T}_{\Delta}=2875$ K is nine times slower than that at 
the equilibrium temperature. Summarizing, the value of $b=0.1$ is a {}``realistic
value'' for the parameter which is responsible for the $J$-dependent
relaxation. 

It is timely to present an exact expression for the angular momentum
integral relaxation time, which is valid for any values of the parameters
of the generalized J-diffusion model. According to the formulas derived
in Appendix B, 

\begin{equation}
\tau_{J}=\frac{1}{\nu}\frac{\xi\eta}{\sqrt{(\eta-b)(\eta\xi^{2}-b)}}\frac{v(b)}{v(0)}\exp\left\{ \frac{\eta bE_{\Delta}}{(\eta-b)}\right\} \label{TjExact}\end{equation}
($v(b)$ is defined via Eq. (\ref{vb})). If we assume that $b\ll\eta,\,\eta\xi^{2}$,
then Eq. (\ref{TjExact}) reduces to (\ref{TauAll}) for any 
$\eta,\,\xi$, and $\Delta$. This lends an additional support
to the validity of Eqs. (\ref{TauAll}). 

On the other hand, Eq. (\ref{TjExact}) demonstrates that the $J$-dependence
of rotational relaxation rates works both ways: if $\Delta$ is
small and $\eta<1$, then nonequilibrium integral relaxation times
can be smaller then their equilibrium counterparts. This is illustrated
by Fig. 3, which shows the reduced angular momentum relaxation time
$\tau_{J}/\tau_{J}^{eq}$ (full line) and rotational energy relaxation
times $\tau_{S}/\tau_{S}^{eq}$ (dashed line) vs. parameter $\eta$.
The nonequilibrium relaxation times are calculated
via Eqs. (\ref{Gen2})-(\ref{wb}), and their equilibrium counterparts
are given by Eqs. (\ref{TauEq}). The observed decrease of the relaxation
times at $\Delta\ll1$ and $\eta<1$ is not so pronounced as their
exponential increase at $\Delta\gg1$. 

Fig. 4 compares rotational CFs (\ref{CJ})-(\ref{CBE}) calculated
within the standard ($b=0$) and generalized ($b=0.1$) J-diffusion
models. The CFs have been computed as explained in Appendix A. The
realistic values of the model parameters ($\eta=1.1$, $\xi=1$, $\Delta=5.4$,
and $\nu=1$) have been used. The standard J-diffusion predicts all
the CFs to coincide with $\exp\{-\nu t\}$. According to the generalized
J-diffusion model, the CFs relax much slowly, as expected. For the
model parameters chosen, $C_{S}(t)$ and $C_{J}(t)$ are almost indistinguishable,
\cite{FootF} while $C_{E}(t)$ decays slower than the former two
CFs. 

If we assume that the initial distribution is given by delta-function
(\ref{DeltaD}) and $b\ll\eta,\,\eta\xi^{2}$, then the general Eqs.
(\ref{Ronet}) and (\ref{sigD}) simplify to yield the following evolution
of the probability density function:
\begin{equation}
\rho(\mathbf{J},t)=\frac{1}{2\pi}\frac{\delta(J-\Delta)}{J}\exp\{-\nu z(\Delta)t\}+\rho_{B}(\mathbf{J})(1-\exp\{-\nu z(\Delta)t\}).\label{Ronet1}\end{equation}
In an accord with what has been observed in, \cite{ger94,str06,str06a}
the rotational distribution (\ref{Ronet1}) is clearly bimodal. At
every time moment, $\rho(\mathbf{J},t)$ is a mixture of the initial
nonequilibrium distribution (\ref{DeltaD}) and the equilibrium distribution
(\ref{Boltz}). The standard J-diffusion model predicts the same formula
but with $z(\Delta)=1$. Thus, the $J$-dependent rate effectively
slows the rotational relaxation and increases the lifetime of the
initial nonequilibrium contribution. The higher is the value of 
$bE_{\Delta}$, the stronger is the effect. For example,
if we take the dimensionless values of the parameters $\nu=0.84$,
$b=0.094$ obtained above, we can calculate the decay rates 
at $\overline{T}_{\Delta}=2875$ K. Namely, the generalized J-diffusion
model predicts $\nu z(\Delta)=0.3$ $\mathrm{ps^{-1}}$, while the
J-diffusion model yields $\nu=2.6$ $\mathrm{ps^{-1}}$. The generalized
J-diffusion rate explains why $\rho(\mathbf{J},t)$ simulated for
CN fragments at $\overline{T}_{\Delta}=2875$ K shows both the nonequilibrium
and equilibrium contributions for more than $4$ ps. \cite{str06a}
The standard J-diffusion model, which predicts the decay rate to be $9$
times higher, is thus absolutely inadequate far from equilibrium. 


\section{Orientational relaxation}

Orientational CF of the rank $j$ is defined through the Wigner D-functions
\cite{var89} as follows:

\begin{equation}
G^{j}(t)\equiv<D^{j}(\Omega(t))D^{j}(\Omega(0))^{-1}>.\label{OCF}\end{equation}
Its Laplace image, $\widetilde{G}^{j}(s)$, can be calculated as
explained in Appendix C, Eq. (\ref{Ocf_fin}). If we take 
the initial delta-function distribution (\ref{DeltaD}), then 
Eq. (\ref{Ocf_fin}) can considerably be simplified. At short times
($t<(\nu z(\Delta))^{-1}$), it can be inverted into the time domain
to yield \begin{equation}
G^{j}(t)\approx\exp\{-\nu z(\Delta)t\}\left(a_{j0}+2\sum_{k=1}^{j}a_{jk}\cos(k\Delta t)\right),\label{ocfgg2}\end{equation}
the numerical coefficients $a_{jk}$ are given by Eq. (\ref{free}).
The standard J-diffusion predicts the same formula but with $z(\Delta)=1$.
Thus, the oscillations in orientational CFs of hot photofragments,
which have been measured in the gas phase \cite{zew94,zew01} and
simulated in the condensed phase, \cite{str06,str06a,wil89,gel00}
are entirely determined by the initial nonequilibrium distribution
(\ref{DeltaD}). The period of these oscillations is
uniquely determined by the value of $\Delta$. 

Fig. 5 compares orientational CFs simulated for hot CN fragments injected
at the temperature $\overline{T}_{\Delta}=2875$ K into the heat bath
of argon atoms at $T=120$ K (Refs. \cite{str06,str06a}) and those calculated
within the generalized J-diffusion model. The delta-function initial
distribution (\ref{DeltaD}) has been used in our calculations, since
it corresponds to the procedure of the preparation of the ensemble of
photofragments in the simulations. The simulated orientational CFs
look qualitatively very similar to those described by Eq. (\ref{ocfgg2}).
Quantitatively, they can be fitted by Eq. (\ref{ocfgg2}) relatively
well, but the so-obtained decay rate of the second rank orientational CF turns
out to be $1.4$ higher than that of the first rank CF. The orientational
CFs which are exactly calculated within the generalized J-diffusion
model are seen to reproduce the simulated CFs very well.
This is remarkable, since we did not fit the simulated CFs: we
used the parameters of the generalized J-diffusion model obtained
in the previous Section through the comparison of the simulated and theoretical
energy relaxation times. Namely, we took $b=0.094$, $\nu=0.84$,
and $\Delta=6.75$. \cite{foot4} This lends an additional
support to self-consistency and predictive strength of the generalized
J-diffusion model. 

Eq. (\ref{ocfgg2}) makes it clear that the persistence of the oscillatory
behavior for orientational CFs is much higher in the generalized J-diffusion
model, due to the $\Delta$-induced decrease of the relaxation rate,
$\nu z(\Delta)$. This is vividly illustrated by Fig. 6, in which
presented are the first and second rank orientational CFs calculated
within the generalized ($b=0.1$) and standard ($b=0$) J-diffusion
models for the {}``realistic values'' of the model parameters ($\eta=1.1$,
$\xi=1$, $\Delta=5.4$, and) and $\nu=7$ . The generalized J-diffusion
model predicts the highly oscillatory orientational CFs. In
exactly the same situation, the standard J-diffusion model predicts overdamped
and slowly decaying CFs, which have nothing in common
with the {}``true'' CFs. Thus, the use of the standard (equilibrium)
models of rotational relaxation beyond their domain of validity may
lead to completely wrong predictions. 

We can use Eq. (\ref{Ocf_fin}) to derive a simple expression for
orientational CFs in the case of strong dissipation, $\nu\gg1$. As
expected, they are described by the diffusion formula
\begin{equation}
G^{j}(t)=\exp\{-Dj(j+1)t\}\label{dif}\end{equation}
with the diffusion coefficient $D$ equaled to the \textit{equilibrium}
angular momentum integral relaxation time, $\tau_{J}^{eq}$. The latter
is given by Eq. (\ref{TauEq}), so that

\begin{equation}
D=\tau_{J}^{eq}=\frac{1}{\nu(1-b)^{2}}.\label{D}\end{equation}

The orientational relaxation times are calculated as 

\begin{equation}
\tau_{\Omega}^{j}=\int_{0}^{\infty}dtG^{j}(t)=\widetilde{G}^{j}(0).\label{TauOR}\end{equation}
It is popular to plot these quantities vs.
the angular momentum relaxation times or rates. \cite{BurTe,Ric77}
That is why Figs. 7 display the graphs $\tau_{\Omega}^{1}$ (a)
and $\tau_{\Omega}^{2}$ (b) vs. $\nu$ calculated within the standard
and generalized J-diffusion models. 

To get a better idea about the behavior of these curves, it is insightful
to obtain explicit expressions for $\tau_{\Omega}^{j}$ in 
case of weak and strong dissipation. If $\nu\ll1$, then Eq. (\ref{Ocf_fin})
predicts that $\tau_{\Omega}^{j}$ is inversely proportional
to the rate $\nu$:\begin{equation}
\tau_{\Omega}^{j}=\frac{c_{j}}{\nu}\left(\frac{\xi\eta}{\sqrt{(\eta-b)(\eta\xi^{2}-b)}}\exp\left\{ \frac{\eta bE_{\Delta}}{(\eta-b)}\right\} +\frac{(1+b)c_{j}}{1-c_{j}}\right).\label{ORsm}\end{equation}

Here $c_{j}=a_{j0}/(2j+1)$ and $a_{j0}$ are explicitly given by
Eq. (\ref{free}). \cite{FootD} The standard J-diffusion formula
\cite{McCl77,BurTe} is recovered at $b=0$. Again, if the product
$bE_{\Delta}$ is $\sim1$ or larger, then nonequilibrium initial
conditions manifest themselves in a considerable increase of $\tau_{\Omega}^{2}$
in comparison with the standard J-diffusion predictions at $\nu<1$
(Figs. 7b). 

When $\nu>1$, the situation changes and the standard J-diffusion
model overestimates the actual values of $\tau_{\Omega}^{j}$ (Figs.
7). If $\nu\gg1$ then the diffusion formula (\ref{dif}) holds
and, therefore,\begin{equation}
\tau_{\Omega}^{j}\tau_{J}^{eq}=\frac{1}{j(j+1)}.\label{ORgr}\end{equation}
This expression can be termed as the generalized Hubbard relation.
The standard Hubbard relation, $\tau_{\Omega}^{j}\tau_{J}=[j(j+1)]^{-1}$,
\cite{BurTe,hub72,Ric77} is seen to be significantly off. This discrepancy
is remarkable, since it embodies the breakdown of the
linear response theory. Indeed, the onset of rotational diffusion
occurs after the molecular angular momenta have been thermolized according
to the bath-induced Boltzmann distribution (\ref{Boltz}). Thus the
rotational diffusion equation (\ref{dif}) is a legitimate description
at $t>\tau_{J}$ and the rotational diffusion coefficient is determined
by the \textit{equilibrium} rotational fluctuations and therefore
by the \textit{equilibrium} value of the angular momentum relaxation
time, $\tau_{J}^{eq}$. As has been detailed in Section 4, if the
system starts far from equilibrium, then its angular momentum relaxation
time $\tau_{J}$ can differ dramatically from its equilibrium counterpart,
$\tau_{J}^{eq}$. It is in this case we expect the breakdown of the
linear response Green-Kubo-type formulas for transport coefficients,
which identify the rotational diffusion coefficient $D$ with the
angular momentum relaxation time $\tau_{J}$.

\section{Conclusion}

An adequate description of rotational and translational relaxation
in liquids under nonequilibrium conditions (or within an interval
of characteristic energies which highly exceeds that of the bath thermal
energies) cannot be carried out in terms of conventional Langevin
and Fokker-Planck equations with constant frictions or by master equations
with constant relaxation rates: one has to take into account (linear
and/or angular) velocity dependence of friction. \cite{gri84,str06,str06a,kos06d,kam,mor74,zhu90,rob91,kos07}
The present paper deals with studying nonequilibrium rotational and
orientational relaxation. We have generalized the standard J-diffusion
model by allowing its dissipation rate to be angular momentum dependent
and calculated various nonequilibrium rotational and orientational
CFs. The reaction $\mathrm{ICN}+h\nu\rightarrow\mathrm{CN+I}$ has been selected as
a prototype process which produces highly nonequilibrium hot photofragments.
We have used the results of computer simulations \cite{str06,str06a}
to obtain realistic values of the model parameters and to test our
theoretical predictions. 

We have found that nonequilibrium rotational relaxation rates assume
the form $\nu\exp\left\{ -bE_{\Delta}/(k_{B}T)\right\} $, where $\nu$
is the equilibrium rate, $E_{\Delta}$ is the initial {}``nonequilibrium
energy'', $T$ is the equilibrium temperature of the heat bath, and
$b$ is the dimensionless small parameter fixed for a system under
study. Accordingly, the relaxation times have the Arrhenius energy
dependence, $\nu^{-1}\exp\left\{ bE_{\Delta}/(k_{B}T)\right\} $.
So, there exist characteristic {}``nonequilibrium energy''
$\overline{E}_{\Delta}\sim k_{B}T/b$ and {}``nonequilibrium temperature''
$\overline{T}_{\Delta}\sim T/b$, starting from which the $J$-dependence
of relaxation rate becomes important and induces a significant slowing
of rotational relaxation. In agreement with simulations,
\cite{str06,str06a} relaxation of the rotational probability density
is shown to be a bimodal time-dependent mixture of the initial nonequilibrium
and final equilibrium distributions. The slowing of rotational relaxation
induces a long lifetime of the initial nonequilibrium contribution,
which manifests itself in pronounced coherent effects. 

The slowing of rotational relaxation causes qualitative
changes in molecular reorientation. In agreement with 
simulations, \cite{str06,str06a} hot nonequilibrium molecules
exhibit slightly perturbed coherent rotation, which persists for several
rotational periods. A similar observation has recently been made in
the context of molecular excitation by strong femtosecond pulses:
nonequilibrium molecular wave packets can be formed in such a way
as to slow their subsequent rotational relaxation. \cite{sei06} 

We have demonstrated that orientational CFs in the overdamped limit
are described by the diffusion equation with the diffusion coefficient
equaled to the \textit{equilibrium} angular momentum integral relaxation
time, $\tau_{J}^{eq}$. Thus the orientational relaxation times $\tau_{\Omega}^{j}$
obey the generalized Hubbard relation, $\tau_{\Omega}^{j}\tau_{J}^{eq}=[j(j+1)]^{-1}$.
Since the nonequilibrium angular momentum relaxation time $\tau_{J}$
can differ substantially from its equilibrium counterpart $\tau_{J}^{eq}$,
the standard Hubbard relation, \cite{BurTe,hub72,Ric77} in which
$\tau_{J}^{eq}$ is replaced by $\tau_{J}$, can be significantly
off. 

It should be emphasized that all rotational and orientational CFs
and their relaxation times dependent explicitly on the nonequilibrium
preparation of the molecular ensemble. This effect is unreproducible within
the standard (equilibrium) rotational models, which predict that 
relaxation rates are independent of initial conditions and are
given by the linear response theory. Thus, the Arrhenius forms of
the rotational relaxation times, slowing down of rotational relaxation,
and violation of the Hubbard relations are all manifestations of the
breakdown of the linear response theory far from equilibrium.

In practical terms, the standard and generalized J-diffusion models
are almost indistinguishable under equilibrium conditions. Under nonequilibrium
conditions, their predictions differ dramatically. The differences
are not only quantitative but, not infrequently, qualitative. The
message is thus as follows: the friction and/or the relaxation rate
must be taken $J$-dependent in order to adequately describe rotational
relaxation far from equilibrium. 

Note, finally, that our theory can straightforwardly be generalized
to symmetric and asymmetric top molecules. In this latter case, the
relaxation rate can be taken rotational energy dependent, rather than
angular momentum dependent. The method of solution of the corresponding
kinetic equations remains absolutely the same, and the explicit formulas
for various CFs can be written down after a straightforward generalization
of the results presented in the Appendixes. Our theory is also readily
extendable to quantum case, by replacing the integrations over
the angular momentum $J$ with summations over the rotational quantum
number $j$, and switching from the classical orientational CFs (\ref{free})
to their quantum mechanical counterparts (see, e.g., Ref. \cite{gel02}).
Such a generalization might be useful for describing rotational coherences
and time-dependent alignments in nonequilibrium dissipative systems,
like those studied in Ref. \cite{sei06}

\appendix

\section{Calculation of rotational CFs}

Let $A(\mathbf{J})$ and $B(\mathbf{J})$ be arbitrary functions of
the angular momentum. Then the CF\begin{equation}
C_{AB}(t)=\left\langle A(\mathbf{J}(t))B(\mathbf{J})\right\rangle _{ne}\label{Cab}\end{equation}
 can be evaluated through the rotational kinetic equation (\ref{kinJ})
as follows:\begin{equation}
C_{AB}(t)=\int d\mathbf{J}A(\mathbf{J})\rho(\mathbf{J},t),\,\,\,\rho(\mathbf{J},0)=\rho_{ne}(\mathbf{J})B(\mathbf{J}).\label{kinjt0}\end{equation}
After applying the Laplace transform (\ref{Lap}) to Eq. (\ref{kinJ})
with initial condition (\ref{kinjt0}), we obtain\begin{equation}
\widetilde{\rho}(\mathbf{J},s)=\frac{\rho_{ne}(\mathbf{J})B(\mathbf{J})+\nu(1+b)z(J)\rho_{B}(J)\widetilde{\sigma}(s)}{s+\nu z(J)}.\label{kinJL1}\end{equation}
Here \begin{equation}
\widetilde{\sigma}(s)\equiv\int d\mathbf{J}z(J)\widetilde{\rho}(\mathbf{J},s).\label{kinJLs}\end{equation}
Multiplying Eq. (\ref{kinjt0}) by $z(J)$ and integrating over $\mathbf{J}$,
we obtain an algebraic equation for $\widetilde{\sigma}(s)$. Inserting
its solution into Eq. (\ref{kinJL1}), multiplying the so obtained expression
by $A(\mathbf{J})$ and integrating over $\mathbf{J}$, we get\begin{equation}
\widetilde{C}_{AB}(s)=\widetilde{\Phi}_{ne}(AB,s)+\nu(1+b)\frac{\widetilde{\Phi}_{B}(zA,s)\widetilde{\Phi}_{ne}(zB,s)}{1-\nu(1+b)\widetilde{\Phi}_{B}(z^{2},s)}.\label{Cab1}\end{equation}
Here we have introduced the functionals \begin{equation}
\Phi_{a}(Y,t)=\int d\mathbf{J}\rho_{a}(\mathbf{J})\exp\{-\nu z(J)t\} Y(\mathbf{J}),\,\,\,\,\, a=B,\, ne\label{psi}\end{equation}
and their Laplace transforms \begin{equation}
\widetilde{\Phi}_{a}(Y,s)=\int d\mathbf{J}\rho_{a}(\mathbf{J})\frac{Y(\mathbf{J})}{s+\nu z(J)},\,\,\,\,\, a=B,\, ne\label{psi1}\end{equation}
which are defined for any function $Y(\mathbf{J})$. 

In principle, Eq. (\ref{Cab1}) delivers the desirable expression
for CF $\widetilde{C}_{AB}(s)$ which, after the numerical inversion
of the Laplace transform, yields the CF in the time domain, $C_{AB}(t)$.
The long-time limit of this CF is determined as follows: \begin{equation}
C_{AB}(t\rightarrow\infty)\equiv\left\langle A\right\rangle _{B}\left\langle B\right\rangle _{ne}.\label{Cabinf}\end{equation}
Here the averages are defined as 

\begin{equation}
\left\langle Y\right\rangle _{a}\equiv\nu\widetilde{\Phi}_{a}(Yz,0)\equiv\int d\mathbf{J}\rho_{a}(\mathbf{J})Y(\mathbf{J}),\,\,\,\, a=B,\, ne.\label{Aev}\end{equation}
If both $\left\langle A\right\rangle _{B}$ and $\left\langle B\right\rangle _{ne}$
are nonzero, then CF (\ref{Cab}) possesses a stationary long-time
asymptote (\ref{Cabinf}). This is so for the rotational energy CF,
for example. Therefore, $\widetilde{C}_{AB}(s)\rightarrow\left\langle A\right\rangle _{B}\left\langle B\right\rangle _{ne}/s$
when $s\rightarrow0$. This is undesirable for doing numerics. It
is more convenient to subtract this constant contribution and redefine
the CF (\ref{Cab}) as follows:\begin{equation}
X_{AB}(t)\equiv\frac{C_{AB}(t)-\left\langle A\right\rangle _{B}\left\langle B\right\rangle _{ne}}{\left\langle AB\right\rangle _{ne}-\left\langle A\right\rangle _{B}\left\langle B\right\rangle _{ne}}.\label{CabNew}\end{equation}
Evidently, this CF is normalized to unity ($X_{AB}(0)=1$) and does
not have any stationary asymptote ($X_{AB}(t)\rightarrow0$ when $t\rightarrow\infty$). 

Taking the Laplace transform of Eq. (\ref{CabNew}), we get\begin{equation}
\widetilde{X}_{AB}(s)\equiv\frac{\widetilde{C}_{AB}(s)-\left\langle A\right\rangle _{B}\left\langle B\right\rangle _{ne}/s}{\left\langle AB\right\rangle _{ne}-\left\langle A\right\rangle _{B}\left\langle B\right\rangle _{ne}}.\label{CabNewS}\end{equation}
Eq. (\ref{CabNewS}) possesses yet an undesirable property: its numerator
is a difference of two terms $\sim1/s$. Of course, these two terms
cancel each other when $s\rightarrow0$ but it 
is convenient  for numerical implementations to explicitly extract this singular contribution
out of $\widetilde{C}_{AB}(s)$. To this end, we make use of the identity

\begin{equation}
\widetilde{\Phi}_{a}(Yz,s)\equiv\frac{1}{\nu}\left\{ \left\langle Y\right\rangle _{a}-s\widetilde{\Phi}_{a}(Y,s)\right\} ,\,\,\,\,\, a=B,\, ne,\label{red}\end{equation}
which is immediately derived from Eq. (\ref{psi1}). Applying this
formula several times, we can rewrite Eq. (\ref{CabNewS})
in the following equivalent form\begin{equation}
\widetilde{X}_{AB}(s)\equiv\frac{\widetilde{\Phi}_{ne}(AB,s)+\left(\nu\widetilde{\Phi}_{B}(z,s)\right)^{-1}\widetilde{\Psi}(s)}{\left\langle AB\right\rangle _{ne}-\left\langle A\right\rangle _{B}\left\langle B\right\rangle _{ne}}.\label{CabNewSF}\end{equation}
Here\begin{equation}
\widetilde{\Psi}(s)\equiv\left\langle A\right\rangle _{B}\left\langle B\right\rangle _{ne}\widetilde{\Phi}_{B}(1,s)+s\widetilde{\Phi}_{B}(A,s)\widetilde{\Phi}_{ne}(B,s)-\left\langle A\right\rangle _{B}\widetilde{\Phi}_{ne}(B,s)-\left\langle B\right\rangle _{ne}\widetilde{\Phi}_{B}(A,s).\label{PS}\end{equation}

Eqs. (\ref{CabNewSF}) and (\ref{PS}) are our final singularity-free
formulas, which are used for the numerical inversion of the Laplace
transforms. The explicit formulas for CFs (\ref{CJ})-(\ref{CBE})
can be obtained by inserting the corresponding functions $A(\mathbf{J})$
and $B(\mathbf{J})$ into Eqs. (\ref{CabNewSF}) and (\ref{PS}).
For example, the angular momentum CF (\ref{CJ}) corresponds to $A(\mathbf{J})=B(\mathbf{J})=\mathbf{J}$.
Due to the isotropy of rotational space, $\widetilde{\Phi}_{a}(\mathbf{J}z^{n}(J),s)\equiv0$,
so that the general expression (\ref{CabNewSF}) simplifies considerably: 

\begin{equation}
\widetilde{C}_{J}(s)=\frac{\widetilde{\Phi}_{ne}(E,s)}{\left\langle E\right\rangle _{ne}}.\label{CJ1}\end{equation}
There is no such a simplification for CFs (\ref{Sne}) and (\ref{CBE}),
and all the terms (\ref{PS}) contribute into Eq. (\ref{CabNewSF}).
We do not give the corresponding expressions explicitly, since they
add nothing profound. 

CF (\ref{CabNew}) can be used to define an important quantity, the
integral relaxation time:\begin{equation}
\tau_{AB}\equiv\int_{0}^{\infty}dtX_{AB}(t)\equiv\widetilde{X}_{AB}(0).\label{TauAB}\end{equation}
Since $\nu\widetilde{\Phi}_{B}(z,0)\equiv1$, Eqs. (\ref{CabNewSF})
and (\ref{PS}) predict that \begin{equation}
\tau_{AB}\equiv\frac{\widetilde{\Phi}_{ne}(AB,0)+\widetilde{\Psi}(0)}{\left\langle AB\right\rangle _{ne}-\left\langle A\right\rangle _{B}\left\langle B\right\rangle _{ne}}.\label{Tab}\end{equation}
Here

\begin{equation}
\widetilde{\Psi}(0)\equiv\left\langle A\right\rangle _{B}\left\langle B\right\rangle _{ne}\widetilde{\Phi}_{B}(1,0)-\left\langle A\right\rangle _{B}\widetilde{\Phi}_{ne}(B,0)-\left\langle B\right\rangle _{ne}\widetilde{\Phi}_{B}(A,0).\label{PS1}\end{equation}

If we assume that the relaxation rate is constant ($b=0,$ $z(J)=1$)
then the J-diffusion model is recovered. In this case, evidently,
\[
\widetilde{\Phi}_{a}(Y,s)=\frac{\left\langle Y\right\rangle _{a}}{s+\nu},\,\,\,\,\widetilde{\Phi}_{a}(Y,0)=\frac{\left\langle Y\right\rangle _{a}}{\nu},\,\,\,\, a=B,\, ne.\]
 Then, Eqs. (\ref{CabNewSF}) and (\ref{Tab}) simplify to

\begin{equation}
\widetilde{X}_{AB}(s)=\left(\nu+s\right)^{-1},\,\,\, X_{AB}(t)=\exp\{-\nu t\},\,\,\,\tau_{AB}=\left(\nu\right)^{-1}\label{JDcf}\end{equation}
 irrespective of particular forms of $A(\mathbf{J})$, $B(\mathbf{J})$,
and $\rho_{ne}(\mathbf{J})$. Thus all rotational CFs in the J-diffusion
model decay exponentially with the rate $\nu$, and their integral
relaxation times are all equaled to the inverse value of this rate. 

If the relaxation rate is $J$-dependent ($b\neq0$), the situation
is much more complicated. However, the formulas simplify dramatically
if the nonequilibrium distribution is given by the delta-function
(\ref{DeltaD}). If we assume, additionally,
that $b\ll1$ but the product $bE_{\Delta}$ can take any value, we
can write then \[
\widetilde{\Phi}_{B}(Y,s)\approx\frac{\left\langle Y\right\rangle _{B}}{s+\nu},\,\,\,\,\widetilde{\Phi}_{ne}(Y,s)=\frac{\left\langle Y\right\rangle _{ne}}{s+\nu z(\Delta)}.\]
After the insertion of these expressions into Eqs. (\ref{CabNewSF})
and (\ref{PS}), we obtain \begin{equation}
\widetilde{X}_{AB}(s)=\frac{1}{\nu z(\Delta)+s},\,\,\, X_{AB}(t)=\exp\{-\nu z(\Delta)t\},\,\,\,\tau_{AB}=\frac{1}{\nu z(\Delta)}\label{TauD}\end{equation}
for any $A(\mathbf{J})$ and $B(\mathbf{J})$. 

If we wish to follow how the initial nonequilibrium distribution $\rho_{ne}(\mathbf{J})$
transforms in time into the equilibrium one, $\rho_{B}(\mathbf{J})$,
we must solve  kinetic equation
(\ref{kinJ}) with the initial condition
$\rho(\mathbf{J},0)=\rho_{ne}(\mathbf{J})$. The solution is given
by a slightly modified version of Eqs. (\ref{kinJL1}) and (\ref{Cab1}):\begin{equation}
\widetilde{\rho}(\mathbf{J},s)=\frac{\rho_{ne}(\mathbf{J})+\nu(1+b)z(J)\rho_{B}(\mathbf{J})\widetilde{\sigma}_{d}(s)}{s+\nu z(J)},\label{Ronet}\end{equation}
\begin{equation}
\widetilde{\sigma_{d}}(s)=\frac{\widetilde{\Phi}_{ne}(z,s)}{1-\nu(1+b)\widetilde{\Phi}_{B}(z^{2},s)}.\label{sigD}\end{equation}

\section{Calculation of rotational integral relaxation times }

If we employ the nonequilibrium initial distribution (\ref{neMy}),
then the quantities $\widetilde{\Phi}_{a}(Y,0)$ (Eq. (\ref{psi1})) can be evaluated
analytically for any CF of interest in the present paper ($Y=1,\, E,\, E^{2}$).
This means that the integral relaxation times $\tau_{AB}$ can also
be calculated analytically. To this end, convenient is to introduce
the generating function\[
\Upsilon(b)=\int JdJ\rho_{ne}(J)\exp\{ b(J_{x}^{2}+J_{y}^{2})/2\}=\]
\begin{equation}
\frac{\xi\eta}{2\pi}\int dJ_{x}\int dJ_{x}\exp\{-\eta(J_{x}-\Delta)^{2}/2\}\exp\{-\eta(\xi J_{y})^{2}/2\}\exp\{ b(J_{x}^{2}+J_{y}^{2})/2\}.\label{Gen}\end{equation}
Evidently, \begin{equation}
\widetilde{\Phi}_{ne}(E^{N},0)=\frac{1}{\nu}\frac{d^{N}\Upsilon(b)}{db^{N}},\,\,\,\,\widetilde{\Phi}_{B}(E^{N},0)=\widetilde{\Phi}_{ne}(E^{N},0)|_{\xi=\eta=1,\,\Delta=0};\label{Gen1}\end{equation}
\begin{equation}
\left\langle E^{N}\right\rangle _{a}=\nu\widetilde{\Phi}_{a}(E^{N},0)|_{b=0},\,\,\,\, a=B,\, ne.\label{Aev1}\end{equation}
Eq. (\ref{Gen}) is elementary evaluated to yield\begin{equation}
\Upsilon(b)=\frac{\xi\eta}{\sqrt{(\eta-b)(\eta\xi^{2}-b)}}\exp\left\{ \frac{\eta b\Delta^{2}}{2(\eta-b)}\right\} .\label{Gen2}\end{equation}
Differentiating this expression with respect to $b$, we obtain:\begin{equation}
\widetilde{\Phi}_{ne}(1,0)=\frac{1}{\nu}\Upsilon(b),\label{E0}\end{equation}

\begin{equation}
\widetilde{\Phi}_{ne}(E,0)=\frac{1}{\nu}\Upsilon(b)v(b),\label{E1}\end{equation}

\begin{equation}
\widetilde{\Phi}_{ne}(E^{2},0)=\frac{1}{\nu}\Upsilon(b)\left\{ v^{2}(b)+w(b)\right\} .\label{E2}\end{equation}
Here

\begin{equation}
v(b)=\frac{1}{2}\left\{ \frac{1}{\eta-b}+\frac{1}{\eta\xi^{2}-b}+\frac{(\eta\Delta)^{2}}{(\eta-b)^{2}}\right\} ,\label{vb}\end{equation}
\begin{equation}
w(b)=\frac{1}{2}\left\{ \frac{1}{(\eta-b)^{2}}+\frac{1}{(\eta\xi^{2}-b)^{2}}+\frac{2(\eta\Delta)^{2}}{(\eta-b)^{3}}\right\} .\label{wb}\end{equation}
These expressions can be substituted into Eq. (\ref{Tab}) to calculate
various integral relaxation times. \cite{FootB} In general, the so-obtained
expressions are quite cumbersome and are not presented here. However,
they are useful for obtaining simple elucidating formulas in several
particular cases. If $b\ll\eta,\,\eta\xi^{2}$ and $\Delta\gg1$,
then \begin{equation}
\Upsilon(b)\approx\exp\left\{ \frac{b\Delta^{2}}{2}\right\} ,\,\, v(b)\approx\frac{\Delta^{2}}{2},\,\, w(b)\approx\frac{\Delta^{2}}{\eta}.\label{Simp}\end{equation}
If the initial distribution (\ref{neMy}) reduces to the equilibrium one ($\eta=\xi=1$,
$\Delta=0$), we get \begin{equation}
\Upsilon(b)=v(b)=\frac{1}{(1-b)},\,\, w(b)=\frac{1}{(1-b)^{2}}\label{Simp1}\end{equation}
and \cite{FootE}\begin{equation}
\tau_{J}^{eq}=\frac{1}{\nu(1-b)^{2}},\,\,\,\tau_{S}^{eq}=\tau_{E}^{eq}=\frac{1+b^{2}}{\nu(1-b)^{3}}>\tau_{J}^{eq}.\label{TauEq}\end{equation}

It has not escaped our notice that the integral relaxation times $\tau_{AB}$
(\ref{Tab}) diverge for $b\geq\eta$, as predicted by Eqs. (\ref{Gen2})-(\ref{wb}).
Although such situation is far beyond the expected domain of validity
of our model ($b\ll\eta$), there is nothing pathological in the occurrence
of the {}``phase transition'' at $b=\eta$ and all CFs are well
behaved for $b\geq\eta$. This means simply that the $J$-dependence
of relaxation rates slows rotational CFs so dramatically that their
integral relaxation times do not exist.

\section{Calculation of orientational CFs}

Having inserted the definition of orientational CF (\ref{OCF}) into
Eq. (\ref{kin2}), we obtain the equation \[
\partial_{t}G^{j}(\mathbf{J},t)=-i\mathbf{J}\mathbf{L}^{j}G^{j}(\mathbf{J},t)-\nu z(J)G^{j}(\mathbf{J},t)\]
\begin{equation}
+\nu(1+b)z(J)\rho_{B}(\mathbf{J})\int d\mathbf{J}'z(J')G^{j}(\mathbf{J}',t).\label{OCF1}\end{equation}
Here $L_{\alpha}^{j}$ are the matrix elements of the angular momentum
operators $\hat{L}_{\alpha}$ over the D-functions: \cite{var89}

\begin{equation}
(L_{x}^{j})_{kl}\pm i(L_{y}^{j})_{kl}=\delta_{k,l\mp1}\{(j\pm l)(j\mp l+1)\}^{1/2},\,\,(L_{z}^{j})_{kl}=l\delta_{kl};\,\,-j\leq k,l\leq j.\label{Jxyz}\end{equation}
 Eq. (\ref{OCF1}) must be solved with the initial condition \begin{equation}
G^{j}(\mathbf{J},t=0)=\rho_{ne}(\mathbf{J}).\label{OCFin}\end{equation}
Evidently,

\begin{equation}
G^{j}(t)\equiv\int d\mathbf{J}G^{j}(\mathbf{J},t).\label{OCFa}\end{equation}

Orientational CFs are calculated very similarly to rotational CFs
(see Appendix A). First we introduce the free linear rotor orientational
CF \cite{ste69} \begin{equation}
F^{j}(J,t)=a_{j0}+2\sum_{k=1}^{j}a_{jk}\cos\{ kJt\},\,\,\,\, a_{jk}=\left(d_{0k}^{j}(\frac{\pi}{2})\right)^{2},\label{free}\end{equation}
 $d_{km}^{j}(\beta)$ being the reduced Wigner function. \cite{var89}
We further define the functional 

\begin{equation}
Q_{a}^{j}(Y,t)=\int d\mathbf{J}\rho_{a}(\mathbf{J})\exp\{-\nu z(J)t\} Y(\mathbf{J})F^{j}(J,t),\,\,\,\,\, a=B,\, ne\label{Gtlin}\end{equation}
and its Laplace transform\begin{equation}
\widetilde{Q}_{a}^{j}(Y,s)=\int d\mathbf{J}\rho_{a}(\mathbf{J})Y(\mathbf{J})\widetilde{F}^{j}(J,s+\nu z(J))=\label{Gtlin}\end{equation}
\[
\sum_{k=-j}^{j}a_{jk}\int d\mathbf{J}\rho_{a}(\mathbf{J})\frac{Y(\mathbf{J})}{s+\nu z(J)+ikJ},\,\,\,\,\, a=B,\, ne.\]
Then, closely following the derivation of Eq. (\ref{Cab1}), we obtain
the following expression for the Laplace transform of the orientational
CF: \begin{equation}
\widetilde{G}^{j}(s)=\widetilde{Q}_{ne}^{j}(1,s)+\nu(1+b)\frac{\widetilde{Q}_{B}^{j}(z,s)\widetilde{Q}_{ne}^{j}(z,s)}{1-\nu(1+b)\widetilde{Q}_{B}^{j}(z^{2},s)}.\label{Ocf_fin}\end{equation}
Since orientational CFs of dissipative molecules, as distinct from
their bath-free counterparts (\ref{free}) and rotational CFs (Appendix
A), do not possess stationary asymptotes, this formula is well behaved
in the limit $s\rightarrow0$ and can be used for the numerical inversion
of the Laplace images into the time domain. 

\begin{acknowledgments}
The authors are grateful to Guohua Tao and Richard Stratt for useful
discussions. This work was partially supported by the American 
Chemical Society Petroleum Research Fund (44481-G6).
\end{acknowledgments}

\clearpage
\begin{figure}
\includegraphics[keepaspectratio,totalheight=20cm]{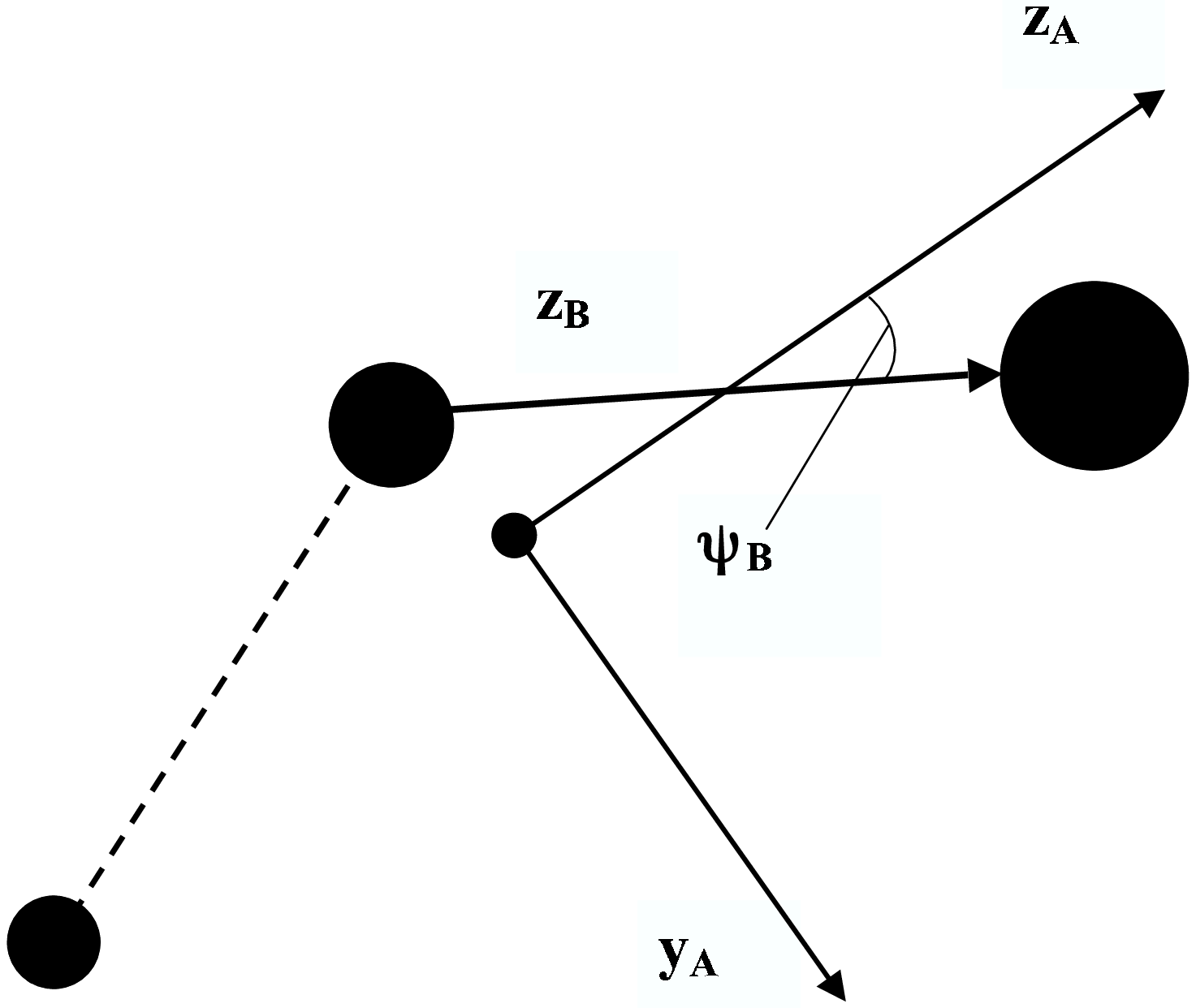}
\caption{Sketch of the dissociating triatomic molecule.}
\end{figure}

\clearpage
\begin{figure}
\includegraphics[keepaspectratio,totalheight=12cm]{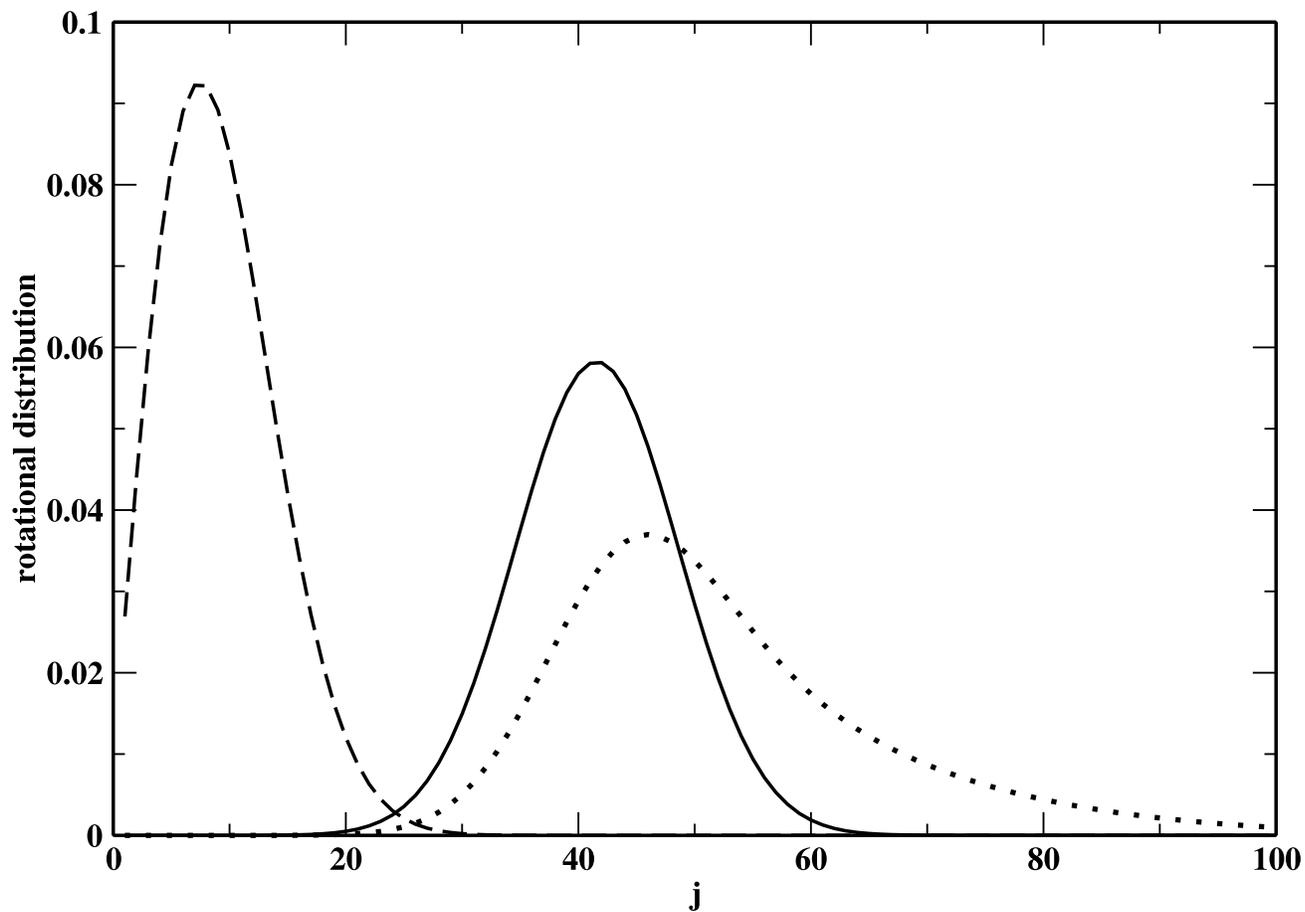}
\caption{Rotational distributions for hot photofragments calculated for $\eta_{Q}=0.02,\,\xi=1$,
$\delta=40$ (full line), $\eta_{Q}=0.02,\,\xi=0.2$, $\delta=40$
(dotted line). The equilibrium Boltzmann distribution for CN at $300$
K is depicted by dashed line ($\eta_{Q}=0.018,\,\xi=1$, $\delta=0$).}
\end{figure}

\clearpage
\begin{figure}
\includegraphics[keepaspectratio,totalheight=12cm]{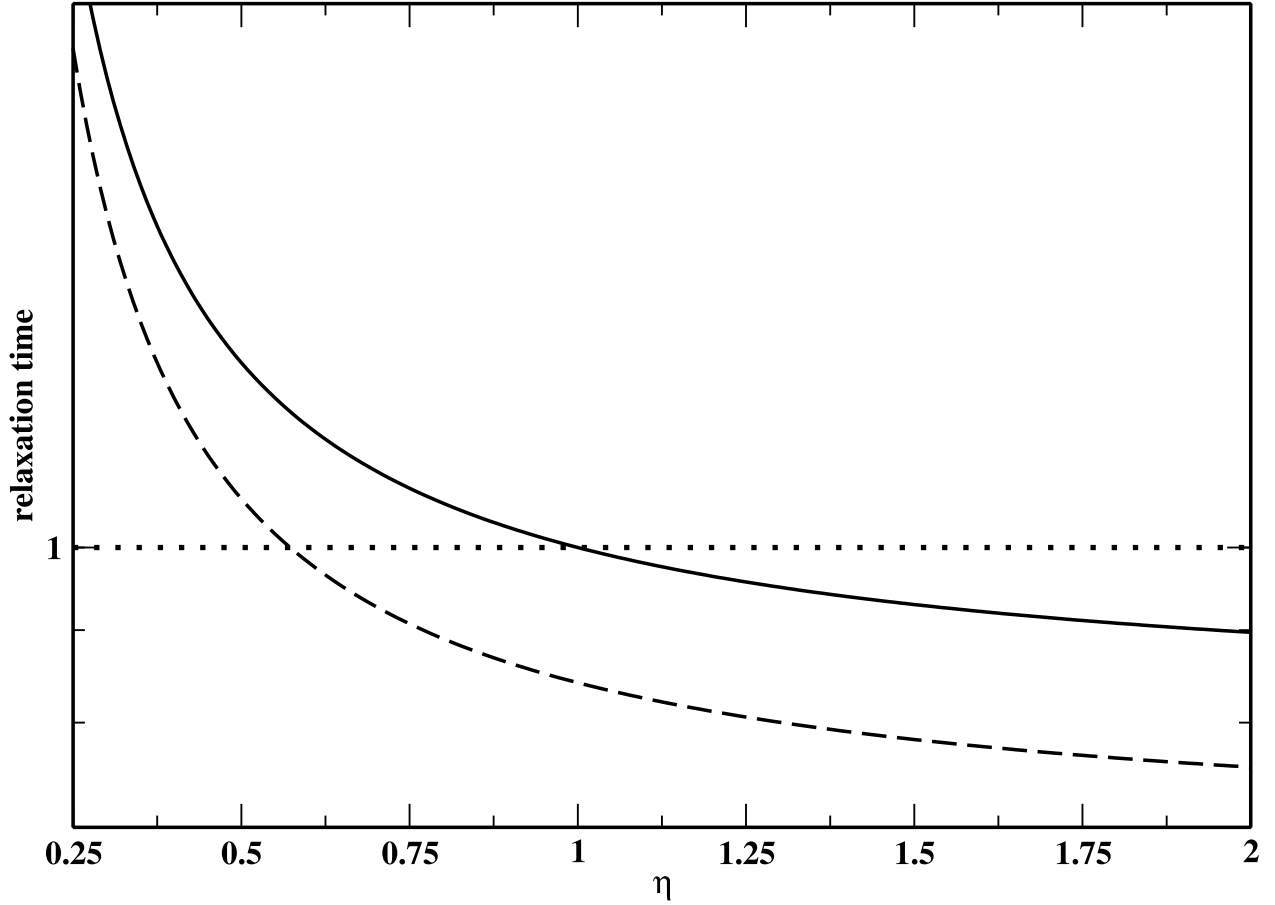}
\caption{Reduced angular momentum relaxation time $\tau_{J}/\tau_{J}^{eq}$
(full line) and rotational energy relaxation time $\tau_{S}/\tau_{S}^{eq}$
(dashed line) vs. parameter $\eta$ calculated in the generalized
J-diffusion model ($b=0.1$) for $\xi=1$ and $\Delta=0$. The dotted
line corresponds to the situation when the reduced value of the relaxation
times equals $1$. }
\end{figure}

\clearpage
\begin{figure}
\includegraphics[keepaspectratio,totalheight=12cm]{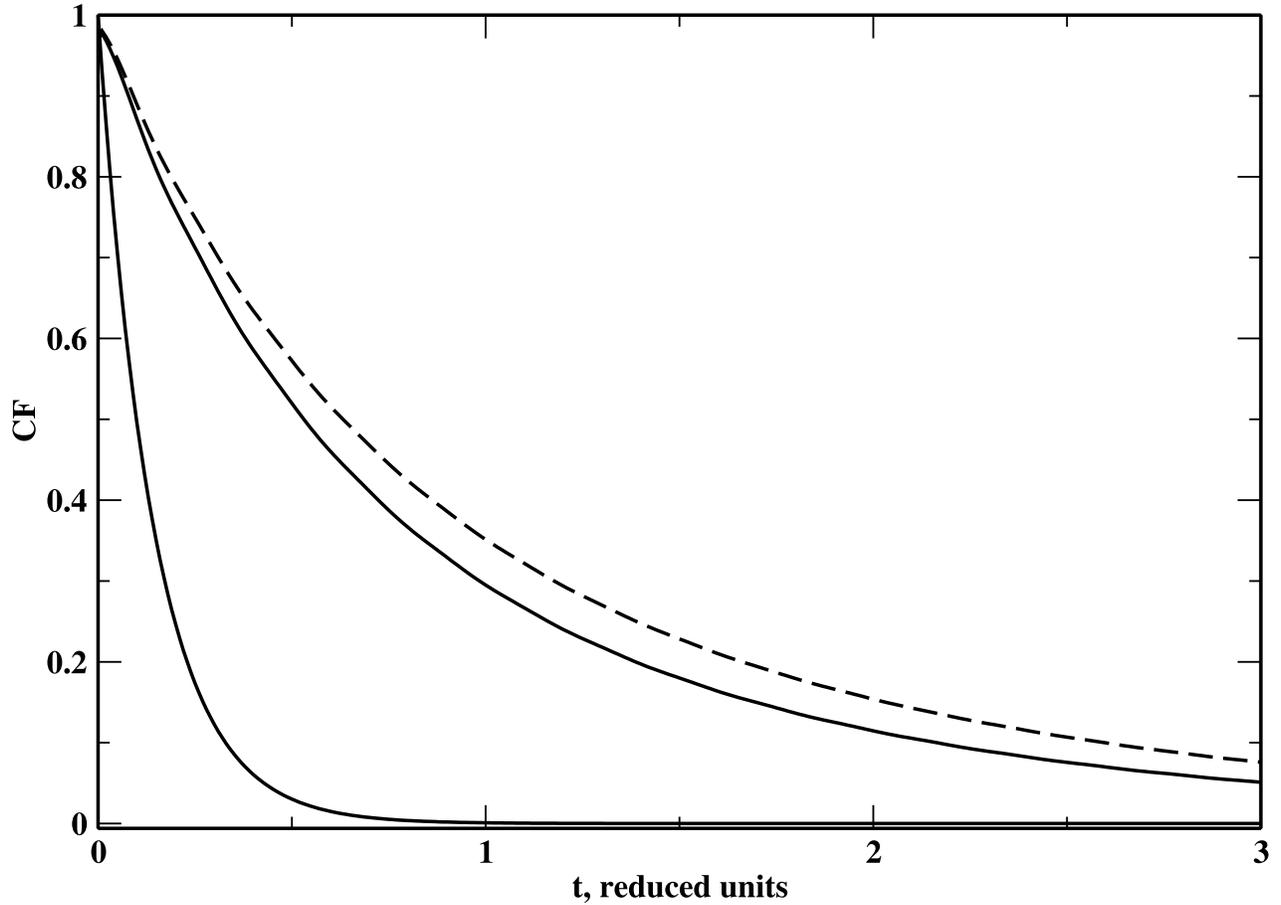}
\caption{ Angular momentum CFs $C_{J}(t)$ (full lines) and rotational
energy CF $C_{E}(t)$ (dashed line) calculated for the {}``realistic
values'' of the model parameters ($\eta=1.1$, $\xi=1$, $\Delta=5.4$)
and $\nu=1$. The lower line corresponds to the standard J-diffusion
model ($b=0$) and the upper lines correspond to the generalized J-diffusion
model ($b=0.1$).  }
\end{figure}

\clearpage
\begin{figure}
\includegraphics[keepaspectratio,totalheight=9cm]{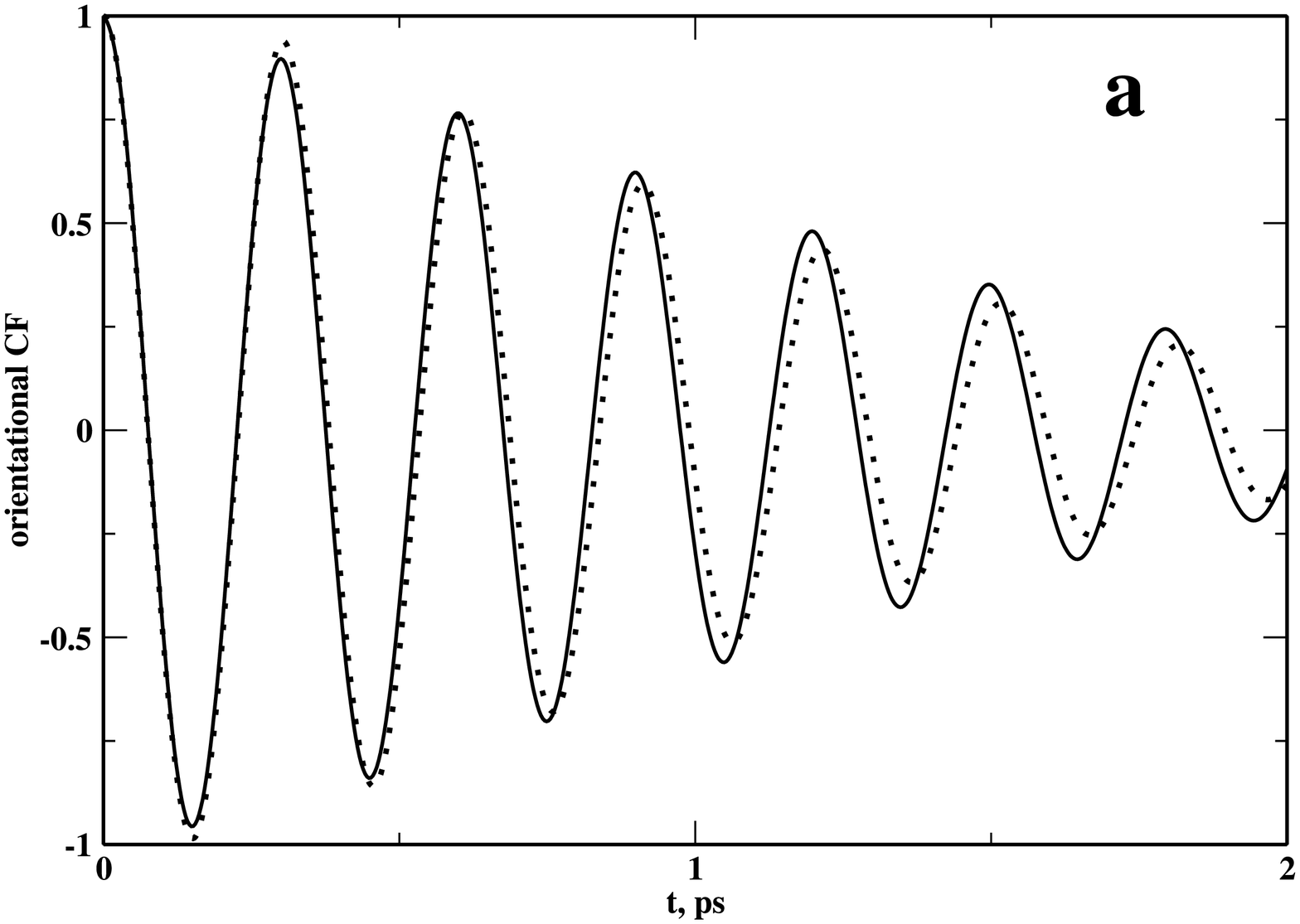}
\includegraphics[keepaspectratio,totalheight=9cm]{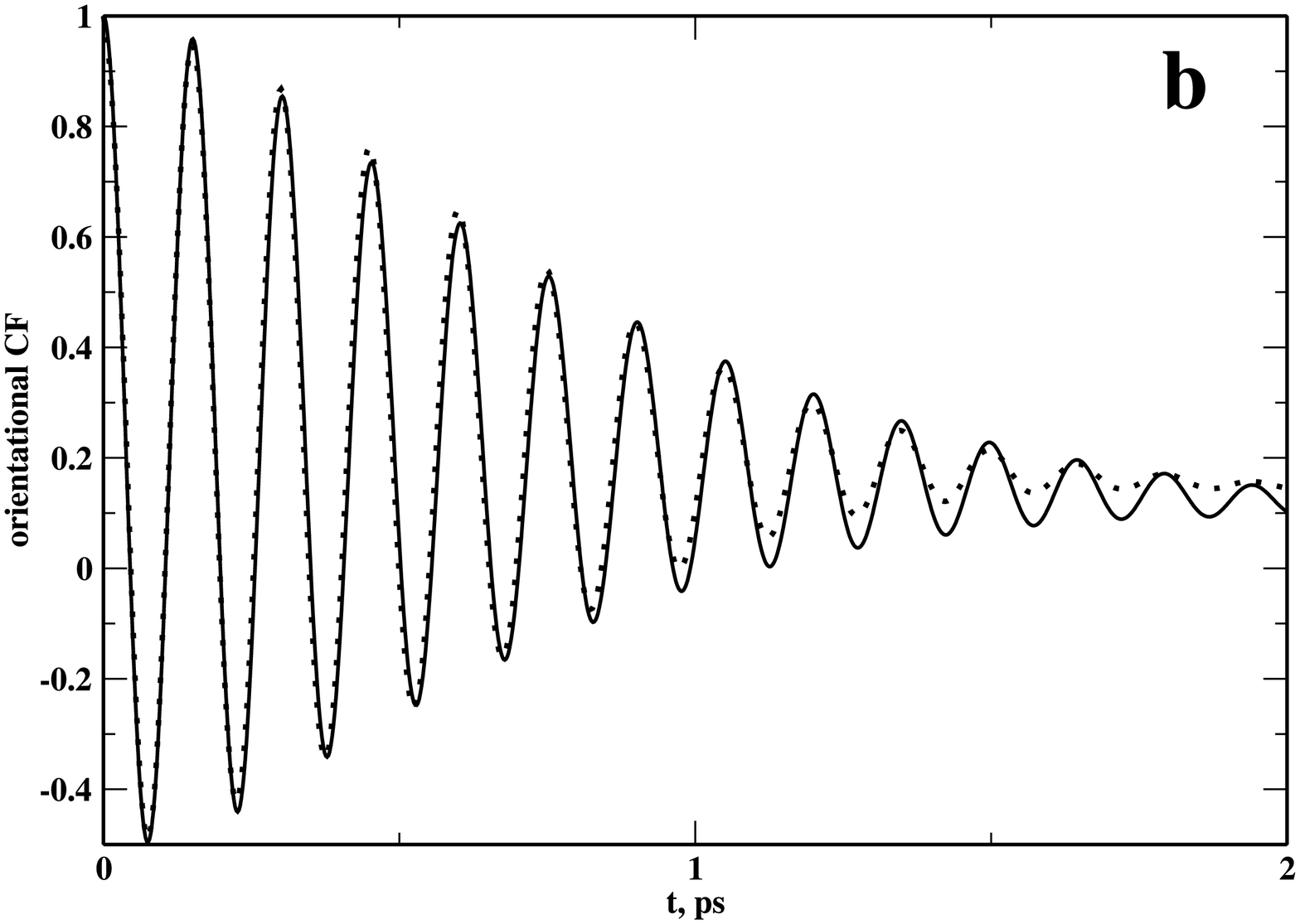}
\caption{ Orientational CFs of the first (a) and second (b) rank. Full
lines depict the results of molecular dynamics simulations of
CN fragments injected at the temperature $\overline{T}_{\Delta}=2875$
K into the heat bath of argon atoms at $T=120$ K (Refs. \cite{str06,str06a}),
and dotted lines show the results of the calculations within the generalized
J-diffusion model ($b=0.094$, $\nu=0.84$) with the initial $\delta$-distribution
(\ref{DeltaD}) centered at $\Delta=6.75$. For CN at $120$ K, $\tau_{r}=$$0.32$
ps.  }
\end{figure}

\clearpage
\begin{figure}
\includegraphics[keepaspectratio,totalheight=12cm]{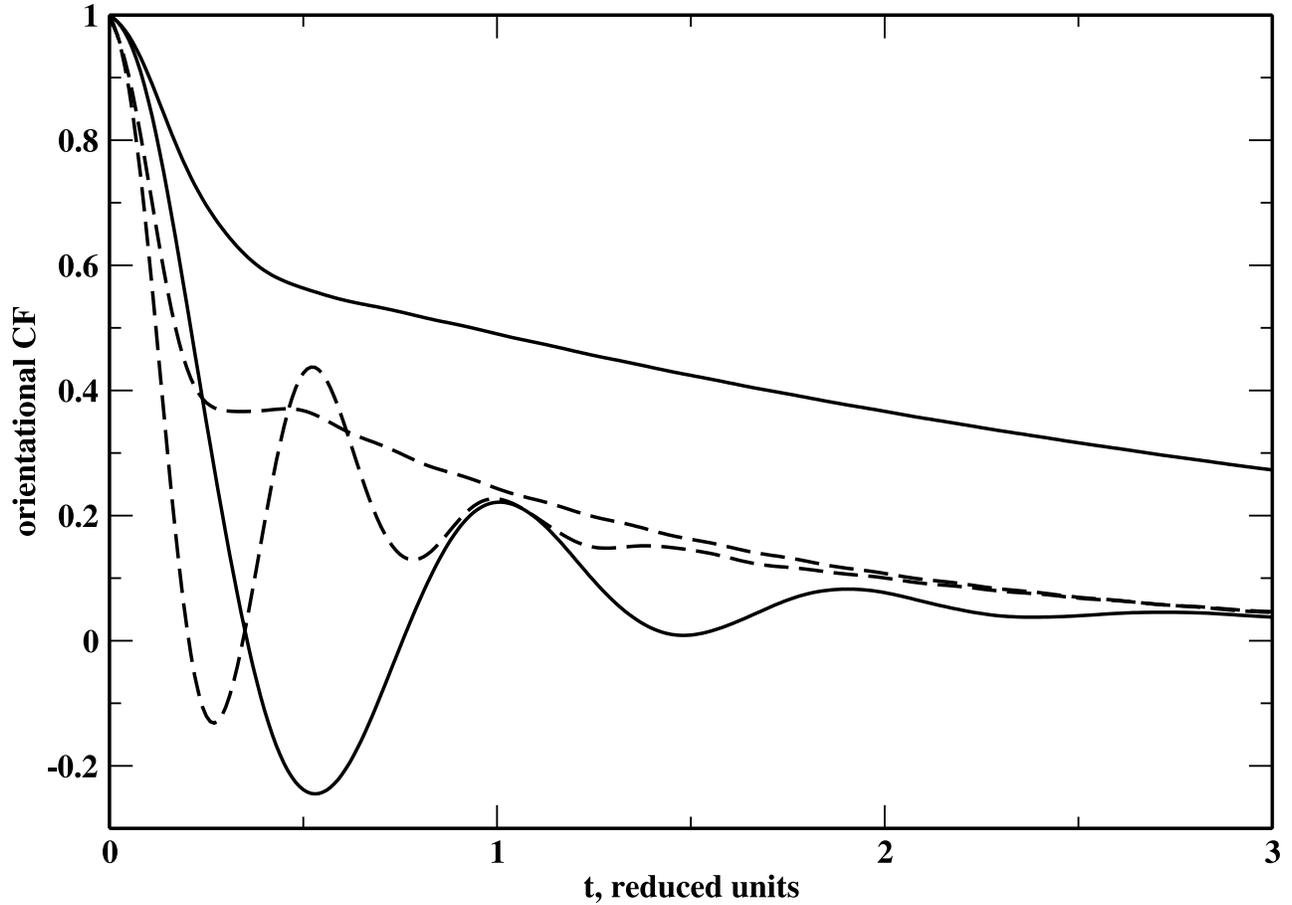}
\caption{ Orientational CFs of the first rank (full lines) and second
rank (dashed lines) calculated for the {}``realistic values'' of
the model parameters ($\eta=1.1$, $\xi=1$, $\Delta=5.4$) and $\nu=7$.
The upper lines correspond to the standard J-diffusion model ($b=0$)
and the lower lines correspond to the generalized J-diffusion model
($b=0.1$). }
\end{figure}

\clearpage
\begin{figure}
\includegraphics[keepaspectratio,totalheight=9cm]{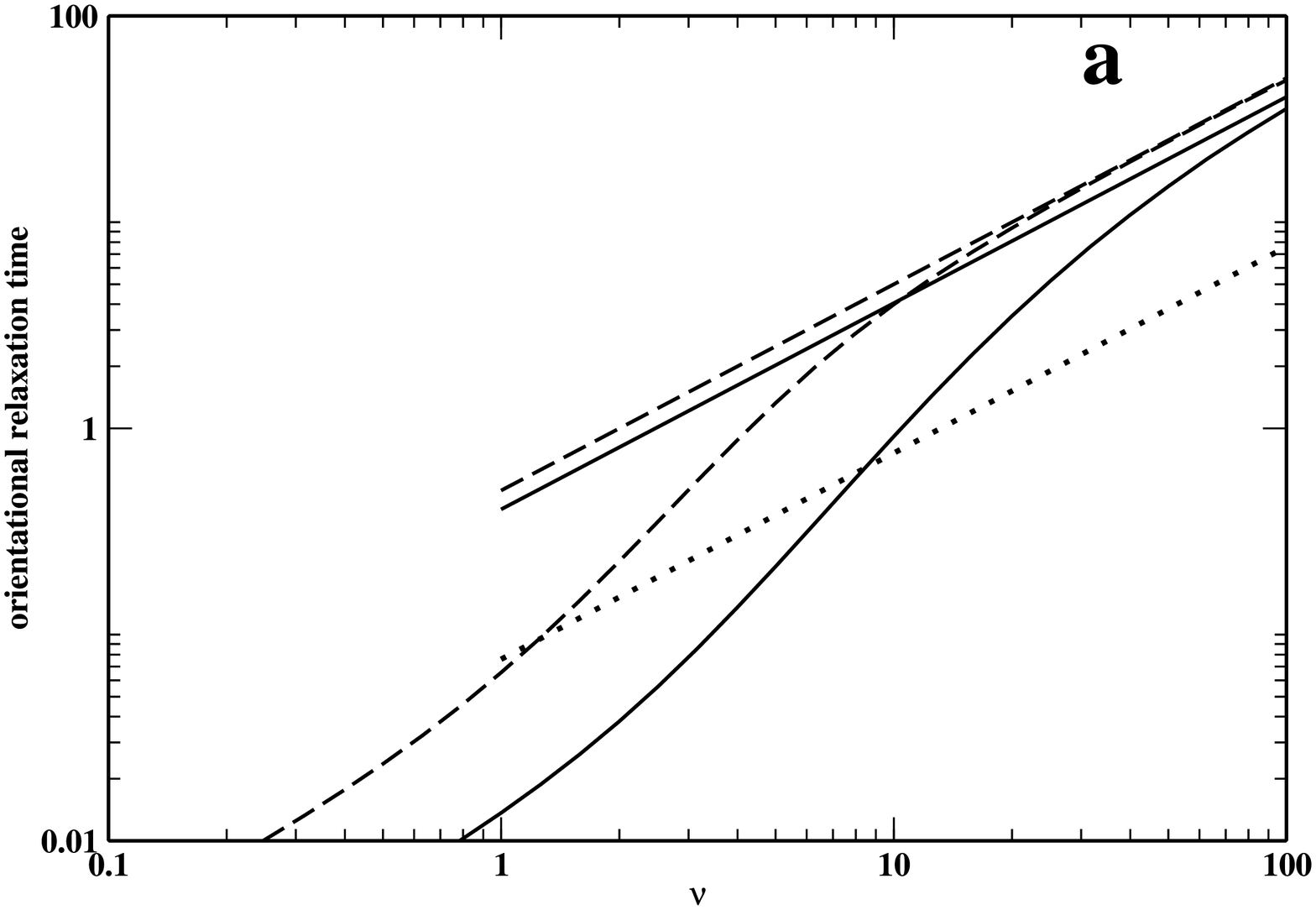}
\includegraphics[keepaspectratio,totalheight=9cm]{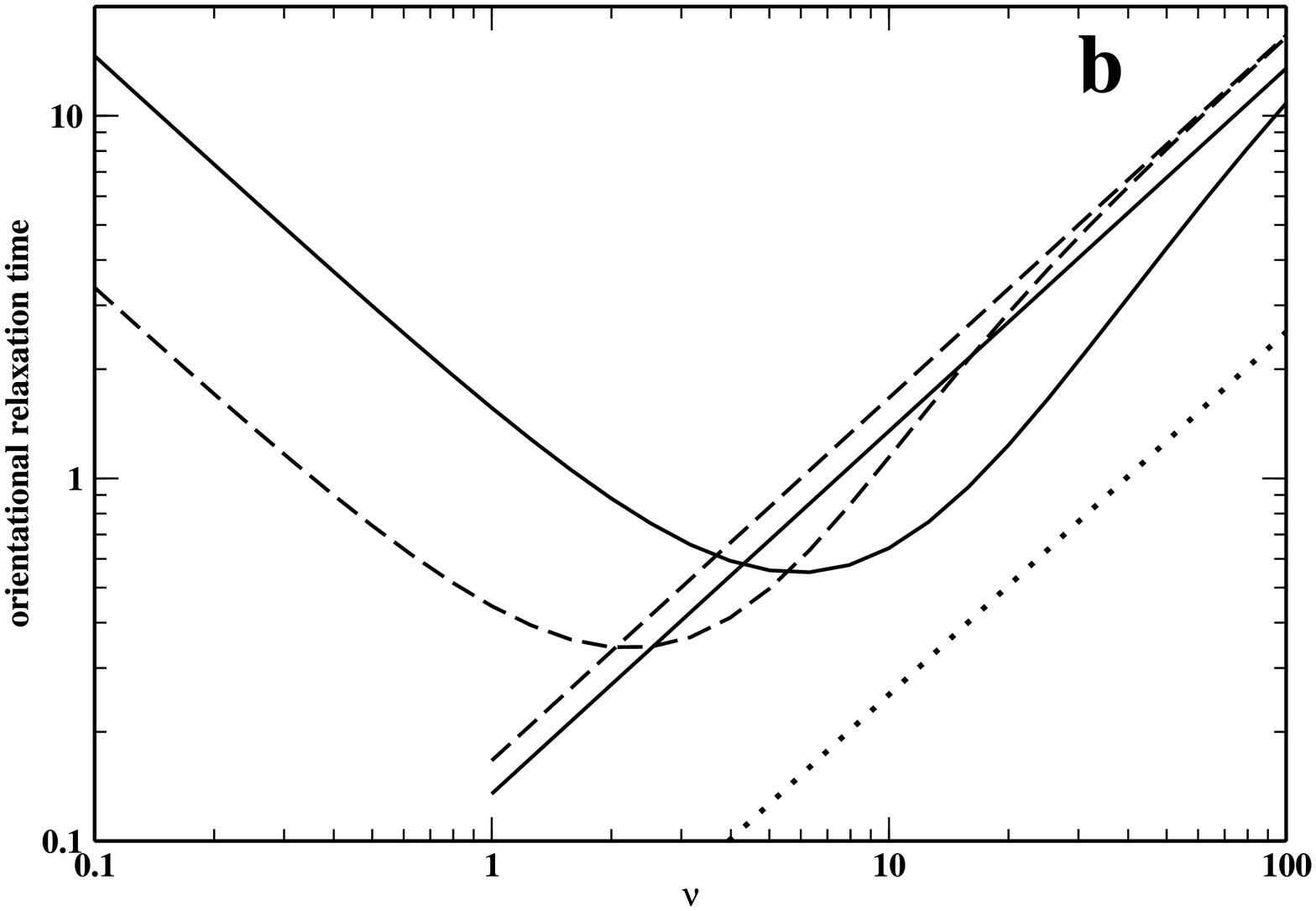}
\caption{ Orientational relaxation times $\tau_{\Omega}^{j}$ of the
first (a) and second (b) rank vs. the collision rate $\nu$ calculated
for the {}``realistic values'' of the model parameters ($\eta=1.1$,
$\xi=1$, $\Delta=5.4$). The dashed lines correspond to the $\tau_{\Omega}^{j}$
and modernized Hubbard asymptote $1/[j(j+1)\tau_{J}^{eq}]$ calculated
within the standard J-diffusion model ($b=0$). The full lines correspond
to their generalized J-diffusion model ($b=0.1$) counterparts. The standard
Hubbard asymptotes, $1/[j(j+1)\tau_{J}],$ are depicted by dotted
lines.   }
\end{figure}


\begin{thebibliography}{20}
\bibitem{FootG}Angular momentum dependent relaxation rates can be calculated
for collisions of rigid nonspherical bodies, \cite{hoa71,fix72,mul93}
but the so-obtained formulas are very cumbersome and the very use
of the predictions of the gas-phase (uncorrelated binary collision)
theories is questionable for our purposes.
\bibitem{foot3}Gaussian expansions are commonly employed in quantum chemical calculations.
We can also recall a similar factorized decomposition of
the spectral density function in quantum relaxation theory. \cite{tan99}
\bibitem{FootA}In some cases, the effect of rotational predissociation on the anisotropy evolution
can be taken into account as proposed in Ref. \cite{gel99}
\bibitem{foot2}Note that the present parameter $\eta$ differs from its counterpart
from Ref. \cite{gel02} by the factor of $2$.
\bibitem{FootF}Note that nonequilibrium CFs $C_{S}(t)$ and $C_{J}(t)$ simulated
in \cite{str06} are also very similar.
\bibitem{foot4}Since the C-N interatomic length has been taken as $1.268$ angstroms
in, \cite{str06,str06a} the free rotation period $\tau_{r}$ for
CN at $120$ K is $0.32$ ps. If we recalculate the nonequilibrium
temperature $\overline{T}_{\Delta}=2875$ into the dimensionless transferred
angular momentum, we get then $\Delta=6.9$. We used a slightly smaller
vale of $\Delta=6.75$ in our generalized J-diffusion calculations
presented in Fig. 5, in order to better reproduce the oscillation
period of the CFs.
\bibitem{FootD}Note that $a_{j0}=c_{j}=0$ for odd $j$. Thus, if $\nu\rightarrow0$,
then also $\tau_{\Omega}^{j}\rightarrow0$ for odd $j$.
\bibitem{FootE}Strictly speaking, $C_{S}(t)$ has  
uncertainty of the kind $0/0$ if we put $ne=B$ in Eq. (\ref{Sne}).
If, however, we consider the limit of $C_{S}(t)$ and $\tau_{S}$
when $ne\rightarrow B$, then the quantities are well behaved.
\end{thebibliography}
\end{document}